\pdfoutput=1
\documentclass[galaxies,article,accept,moreauthors,pdftex]{mdpi} 
\usepackage{multirow}
\externalbibliography{yes}
\usepackage{longtable}
\usepackage{psfrag}
\usepackage{latexsym}
\usepackage{lscape}
\usepackage[export]{adjustbox}
\usepackage{verbatim}
\usepackage{mathrsfs}
\usepackage{bm}
\usepackage{fancyhdr}
\usepackage{multirow}
\usepackage{caption}
\usepackage{subcaption}
\usepackage{booktabs}
\usepackage{xcolor}
\usepackage{rotating}
\usepackage{amsmath,amsfonts,amsthm,bm}

\firstpage{1} 
\pubvolume{1}
\issuenum{1}
\articlenumber{0}
\pubyear{2021}
\copyrightyear{2021}
\externaleditor{Academic Editors: Francesca Loi and
Tiziana Venturi} 
\datereceived{} 
\dateaccepted{} 
\datepublished{} 
\hreflink{https://doi.org/} 

\Title{{Third-Generation} Calibrations for MeerKAT observation}

\TitleCitation{Third-Generation Calibrations for MeerKAT observation}
\AuthorCitation{Parekh, A.; Kincaid, R.; Hugo, B.; Ramaila, A.; Oozeer, N}

\Author{{Viral Parekh} $^{1,2,}$*\orcidA{},  
 Robert Kincaid $^{2}$, Benjamin Hugo $^{1,2}$, Athanaseus Ramaila $^{1}$ and Nadeem Oozeer $^{1, 3,}${*}\orcidA{}}

\address{%
$^{1}$ \quad South African Radio Astronomy Observatory, 2 Fir Street, Black River Park, Observatory, Cape Town 7925, South Africa;  bhugo@ska.ac.za (B.H.); aramaila@ska.ac.za (A.R.) \\
$^{2}$ \quad Department of Physics and Electronics, Rhodes University, PO Box 94, Makhanda 6140, South Africa; kincaidr@ymail.com \\
$^{3}$ \quad African Institute for Mathematical Sciences, 6 Melrose Road, Muizenberg 7945, South Africa\\
}
\corres{Correspondence: {viral.parekh2912@gmail.com} (V.P.); nadeem@ska.ac.za (N.O.)}

\begin{abstract}{Superclusters and galaxy clusters offer a wide range of astrophysical science topics with regards to studying the evolution and distribution of galaxies, intra-cluster magnetization mediums, cosmic ray accelerations and large scale diffuse radio sources all in one observation.  Recent developments in new radio telescopes and advanced calibration software have completely changed data quality that was never possible with old generation telescopes. Hence, radio observations of superclusters are a very promising avenue to gather rich information of a large-scale structure (LSS) and their formation mechanisms. These newer wide-band and wide field-of-view (FOV) observations require state-of-the-art data analysis procedures, including calibration and imaging, in order to provide deep and high dynamic range (DR) images with which to study the diffuse and faint radio emissions in supercluster environments. Sometimes, strong point sources hamper the radio observations and limit the achievement of a high DR. In this paper, we have shown the DR improvements around strong radio sources in the MeerKAT observation of the {Saraswati} supercluster by applying newer third-generation calibration (3GC) techniques using CubiCal and killMS software. We have also calculated the statistical parameters to quantify the improvements around strong radio sources. This analysis advocates for the use of new calibration techniques to maximize the scientific returns from new-generation telescopes.}
    
\end{abstract}
\keyword{diffuse radio sources; radio galaxies; galaxy clusters;  next-generation radio telescopes}
\begin{document}
\label{firstpage}
    

\section{Introduction}

\par The large-scale structure (LSS) of the universe takes the form of an intricate, inter-connected web-like structure called the cosmic web \citep{2014MNRAS.438.3465T}. This structure is a result of hierarchical structure formation processes attributed to the current, widely accepted model of cosmology, the~$\lambda$CDM model. The~LSS is composed of sheet-like filaments and voids, which are continuously evolving due to these hierarchical merging processes taking place at the intersection points or nodes of the filaments \citep{2015JCAP...01..036J,2015A&A...580A.119V,2005Natur.435..629S}. The~largest units, galaxy clusters, are born at these junction points and further evolved through the cluster merging processes. Recent improvements of radio interferometric telescopes, such as MeerKAT, LOFAR, uGMRT, and ASKAP \citep{2018ApJ...856..180C,2013A&A...556A...2V,2017CSci..113..707G,2016PASA...33...42M}, have provided us with the opportunity to conduct high sensitivity, high angular resolution observations of these complex supercluster~objects. 

\par The improved sensitivities of these radio telescopes allow us to study faint and diffuse radio sources such as radio halos, relics, and~mini-halos, which are associated with the whole cluster evolution process and not with any single galaxy \citep[and references therein]{2019SSRv..215...16V}. These radio sources have sizes of a few 100~s kpc to 1 Mpc depending on their host cluster mass and redshift.  Currently, there are more than 100 diffuse radio sources that have been detected in the intra-cluster medium (ICM) of galaxy clusters. Based on simulations and modelling, it has been shown that, generally, radio halos are generated due to the turbulence activated by any form of cluster disturbance process and radio relics are due to cluster merger shocks.  Recently, diffuse radio emissions have also been detected between two clusters in the form of a `radio ridge' \citep{2019Sci...364..981G} and their formation is difficult to explain using standard~theories. 

\par In this paper, we show the importance of the third-generation calibration (3GC, \citep{2010A&A...524A..61N}) on wide field-of-view (FOV) radio interferometry observations, for~example, the~{Saraswati} supercluster~\citep{2017ApJ...844...25B}.  In~radio astronomy, subsequent developments in calibration and imaging techniques have  enhanced our knowledge of the radio universe. Traditionally, 1GC, or first-generation calibration, describes the transfer of complex gain derived from amplitude and phase calibrators to the target source to fix the amplitude and phase variations. In~the 1970s, 2GC, or second-generation calibration, was proposed, which is known as self-calibration~\citep{1984ARA&A..22...97P}. This 2GC uses the target itself to calibrate the phase and amplitude fluctuations. The~2GC  has opened up the field of radio astronomy and provided deep and detailed imaging of radio sources with the highest dynamic range (DR). Both 1GC and 2GC are direction-independent (DI) calibration techniques. However, self-calibration often has limitations in terms of correcting gains of off-centre complex radio sources.
Furthermore, these radio sources are often contaminated by antenna sidelobes, atmospheric fluctuations, telescope pointing inaccuracies and~other unknown factors. To~correct the calibration errors towards these sources, one has to use a new generation of calibration process that can work not only for the phase-centre source(s), but~also improve the sources away from the tracking centre. This requires special software which can break the large FOV into a number of pieces and correct complex gains for the particular patch of the sky. This is also known as direction-dependent effect (DDE or DD) calibration. DD calibration is important for future radio telescopes in order to improve both the DR and image quality \citep{2016ApJS..223....2V,2019A&A...622A...1S,2021A&A...648A...1T}. In~this work, we aim to use DDE techniques for MeerKAT's large FOV observations and quantitatively analyse the observational improvements before and after the DDE using the software {Aimfast} \endnote{\url{https://github.com/Athanaseus/aimfast}} 
. We show improvements around strong radio sources after the DDE, with~several statistical measurements to quantify the observational enhancement. This paper is structured as follows: In Section~\ref{data_ana}, we give brief details of the MeerKAT observations of the {Saraswati} supercluster. In~Section~\ref{3GC}, we describe the 3GC calibration techniques and their applications to the MeerKAT data and finally, Section~\ref{Discussion} is the discussion and conclusion.  

\section{MeerKAT Observations of the Saraswati~Supercluster} \label{data_ana}


\par MeerKAT is a precursor to the square kilometre array (SKA), located in the Karoo Desert in South Africa \citep{2009IEEEP..97.1522J}. The~MeerKAT radio interferometer comprises  64 dish antennas operating at L- and UHF-band. We have conducted the pilot observations of the massive {Saraswati} supercluster with the MeerKAT telescope at L-band. In~these observations, we observed two massive galaxy clusters, A2631 and Zwcl2341.1+0000, for~a total observation time of 14 h (4016 channels and 856 MHz bandwidth). These two clusters are situated near the core of the {Saraswati} supercluster, surrounded by the galaxies' 
filamentary network. The~pilot project aims to map this rich, dense and peculiar supercluster's core using the MeerKAT telescope. The previous observation of ZwCl2341.1+0000, in~NVSS data, has revealed the existence of large-scale diffuse radio emissions in the cluster and later high-resolution Giant Metrewave Radio Telescope (GMRT; 610, 241 and 157 MHz) observations characterised the diffuse emission in the form of double relics~\cite{2002NewA....7..249B, 2009A&A...506.1083V}, which are situated in the north-west and south-east directions, respectively. We have analysed the MeerKAT L-band data with the fully automated CaraCal pipeline~\cite{2020arXiv200602955J}, followed by DDE calibration with killMS \citep{2014arXiv1410.8706T,2014A&A...566A.127T} and CubiCal \citep{2018MNRAS.478.2399K} software. For~imaging and deconvolution purposes, we used the DDFacet software \citep{2018A&A...611A..87T}. We described the full MeerKAT observation details of the {Saraswati} supercluster and its data analysis methods in another paper \citep{2021arXiv211007713P}. In~this paper, we only focus on ZwCl2341.1+0000 to show the elimination of artefacts and improvements around double radio relics and strong radio sources after DDE calibrations using CubiCal and killMS. Our subsequent papers will discuss the science further, specifically the origin of diffuse radio sources in supercluster environments (Kincaid~et~al. in prep.) and radio galaxy properties in the core of the Saraswati supercluster (Parekh~et~al. in prep.)


\section{Third-Generation~Calibrations} \label{3GC}

\par The emergence of newer and more sophisticated radio telescopes have included properties such as non-coplanar arrays, large FOVs and large fractional bandwidths. These have consequently brought a host of new calibration and observational errors that require new and elaborate calibration techniques. Furthermore, detecting such a faint radio emission is extremely hard; sometimes, nearby strong and bright radio sources  limit the ability to observe weak radio signals of astrophysical sources due to artefacts. Even with such sophisticated instruments and state-of-the-art data processing pipelines, astronomical data is still populated with a host of observation- and calibration-related errors. These errors are classified as DDE errors and are source specific. They can be visually recognised by the point spread function (psf) sidelobe-like artefacts they produce around strong sources, most significantly around sources located far away from the phase centre. Sometimes, these strong artefacts, in~radial-spikes form, corrupt the whole image and hamper the achievable DR at the science target position. Accounting for these DDE errors is currently achieved through the use of more modern, third-generation calibration (3GC) techniques. Below, we describe two such methods that are tested and could be implemented in the future MeerKAT data analysis~pipeline.

\subsection{Source Peeling with~CubiCal}
\par CubiCal performs DD calibration using a simultaneous form of the peeling approach called differential gains~\cite{2011A&A...527A.106S}. It can apply corrections to many sources simultaneously from the self-calibrated sky model, as~compared to the one-by-one iterative algorithm used by the typical peeling method \citep{2019RNAAS...3..110W}. These sources are manually marked and contain information on the specific direction in which to perform peeling. The~radio interferometer measurement equation (RIME, \citep{2011A&A...527A.106S,2011A&A...527A.107S}) used for the differential gains takes the form:
\begin{equation} 
D_{l,m} =  G_{l}\bigg(\sum_{l,m} \sum_s \Delta E_{l}P_{l}S_{l,m,s}P_{m}^{H} \Delta E_{m}^{H} \bigg ) G_{m}^{H}
\end{equation}
{${G}$} 
 are DI errors affecting the entire FOV, constrained to phase-only solutions while {$\Delta {E}$} are direction-dependant errors which are notably present around bright sources far away from the phase centre. These errors are constrained to fully complex $2 \times 2$ Jones matrices. $P$ terms are implemented for antenna-specific primary beam models to correct for time and frequency varying gains, and~$S$ terms denote sky models or clean component models of the given observation. The~sum is over all sources $S$ and their respective directions $l,m$.
CubiCal solves for the {${G}$} terms (DI) on small-time/frequency scales by the field as a whole while simultaneously solving for {$\Delta {E}$} terms (DDE) on larger time/frequency scales for a subset of~sources.

\subsection{Facet-Based Calibration with~killMS}
\par The killMS performs DD self-calibration based on a model or image using the Wirtinger formalism applied to the non-linear optimization problem in the context of calibration~\cite{2014arXiv1410.8706T}.
It uses the special linear operator, the~Wirtinger Jacobian, to~perform simultaneous n-directional solving. The killMS runs sequentially over the individual measurement sets and implements two very efficient algorithms for solving the DD calibration problem, namely the Jones based solver: complex half-Jacobian optimization (CohJones) and a physics-based solver, the~non-linear Kalman filter (KAFCA) \cite{2015MNRAS.449.2668S}. For~a given solver, the solution intervals over time and frequency must be specified as a means to achieve a higher signal to noise ratio (SNR). The killMS divides the image plane in a number of tessellations depending on the SNR of the data. Then, various steps in the calibration can be performed in parallel per direction or tessellation. DDFacet does wide-band spectral deconvolution on the tessellated images while taking these Wirtinger DDE solutions (Jones matrices from killMS) into~account.


\subsection{Applications of the DDE to MeerKAT~Data}
\par We have applied CubiCal and killMS software to our MeerKAT data of the {Saraswati} supercluster in order to improve the image quality and reduce calibration-related artefacts and errors. {W{e used standard calibration parameters of killMS (v2.7.0)} and {CubiCal (v1.5.11)} \endnote{\url{https://github.com/cyriltasse/DDFacet/wiki/Creating-a-MeerKAT-DD-corrected-intrinsic-flux-image-with-DDF-kMS} and \url{https://cubical.readthedocs.io/en/latest/examples.html}} which are suitably applicable for MeerKAT data}. In~this paper, we show the results of only one cluster, ZwCl2341.1+0000, before~and after DDE corrections. We have also statistically quantified the improvements after the DDE by using the standard statistical parameters. In~the ZwCl 2341.1+0000 full image (Figure \ref{DDE_corr_1} (top)), we marked five strong sources (A to E) according to their distances (near to far) from the phase centre, which generates artefacts. We peeled these five sources simultaneously using CubiCal and show the DDE corrected image in Figure~\ref{DDE_corr_1} (middle). Similarly, in~killMS, we chose five directions corresponding to these five strong sources. Thereafter, killMS calibrated the data using five facets which covered the whole FOV. Then, each of these facets were calibrated separately to correct for the complex gain solutions. We show the killMS-corrected image in Figure~\ref{DDE_corr_1} (bottom). We have also shown the central zoomed region of ZwCl 2341.1+0000 (corresponds to Figure~\ref{DDE_corr_1}) in Figure~\ref{DDE_corr_2}, where double radio relics are situated. As~can be seen, there are clear improvements after the DDE in the region around the relics and significant artefacts generated by the strong source A were eliminated.

\begin{figure}[H]
\includegraphics[scale=0.35]{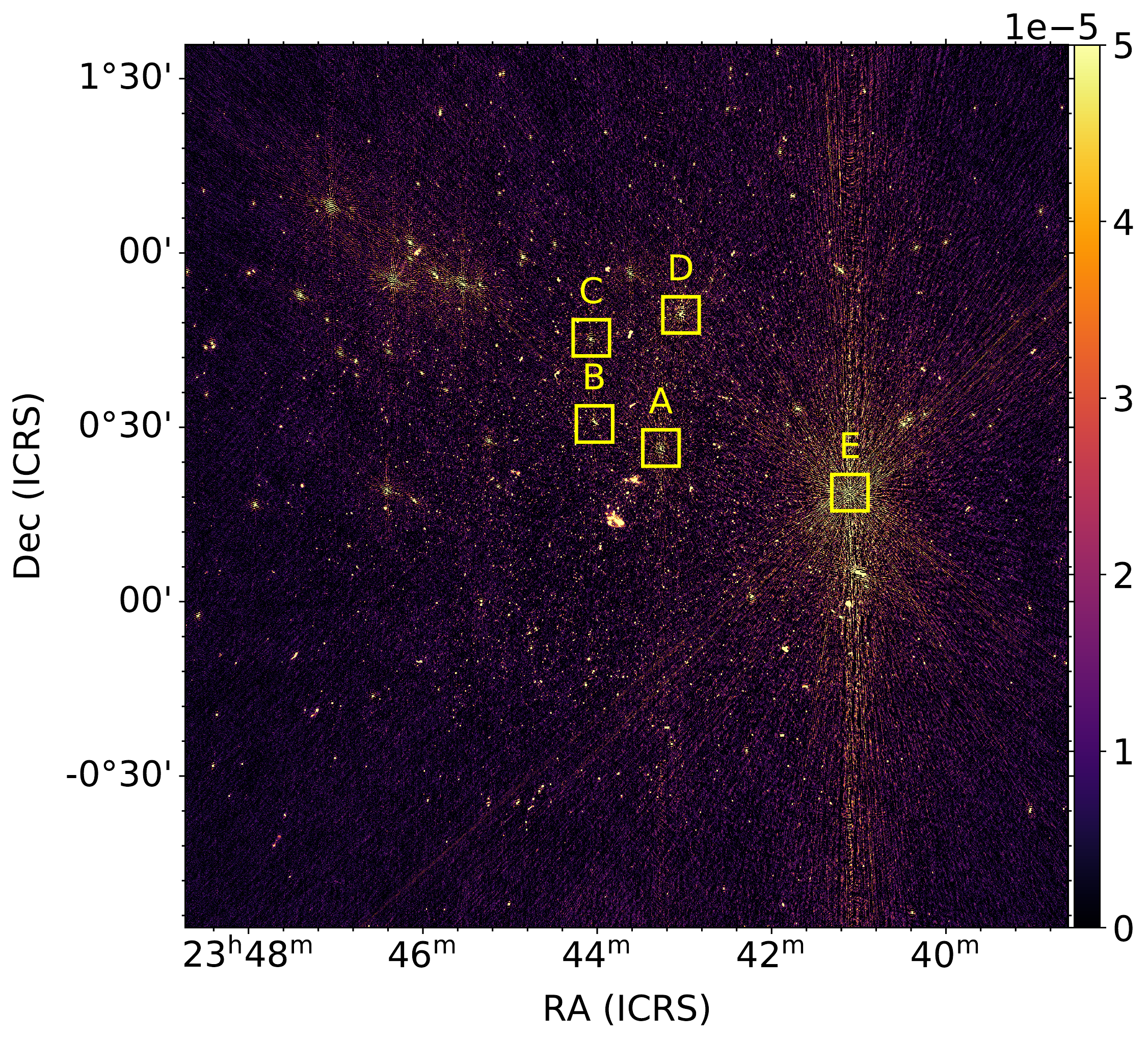}\\
\includegraphics[scale=0.35]{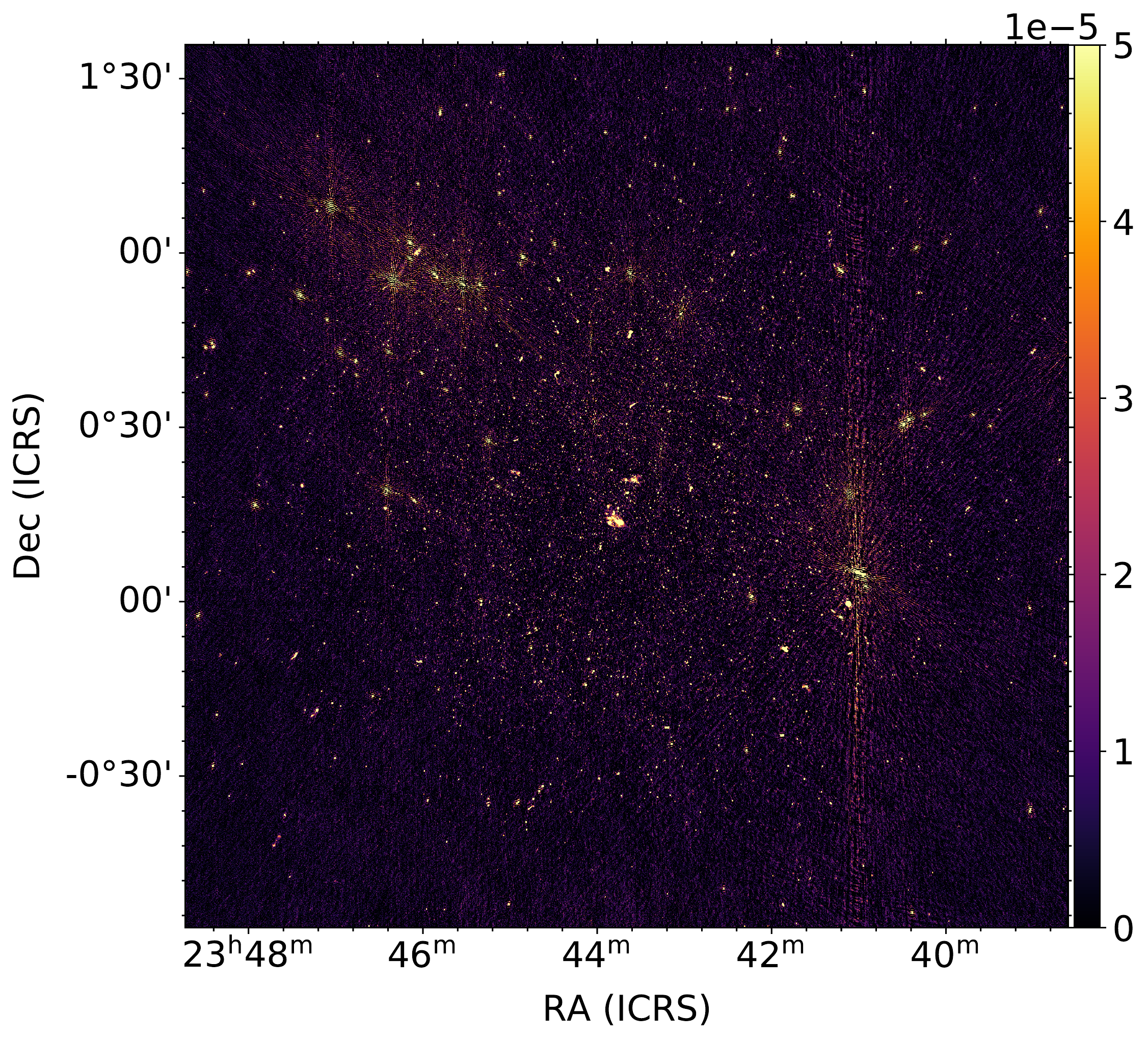}\\
\caption{\textit{Cont}.}
\end{figure}

\begin{figure}[H]\ContinuedFloat
\includegraphics[scale=0.35]{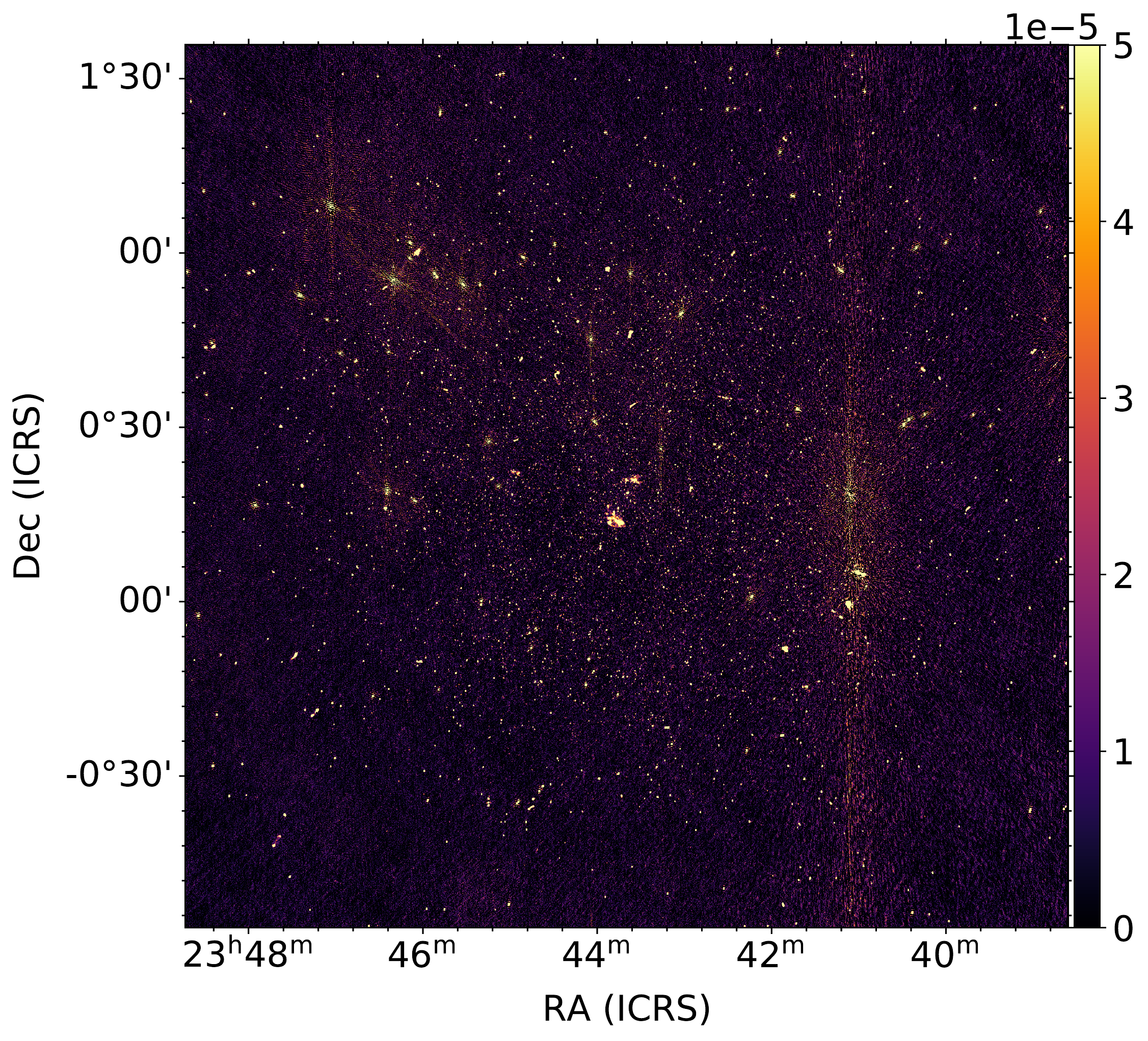}
\caption{{ZwCl 2341.1+0000} full image. (Top) DI image. Five strong sources are marked in boxes with A-E alphabets. (Middle) DD image generated with the CubiCal. (Bottom) DD image generated with the killMS. The~beam sizes are 8$''$ $\times$ 6$''$ for all~images.}
\label{DDE_corr_1}
\end{figure}

\vspace{-6pt}

\begin{figure}[H]

    \includegraphics[scale=0.33]{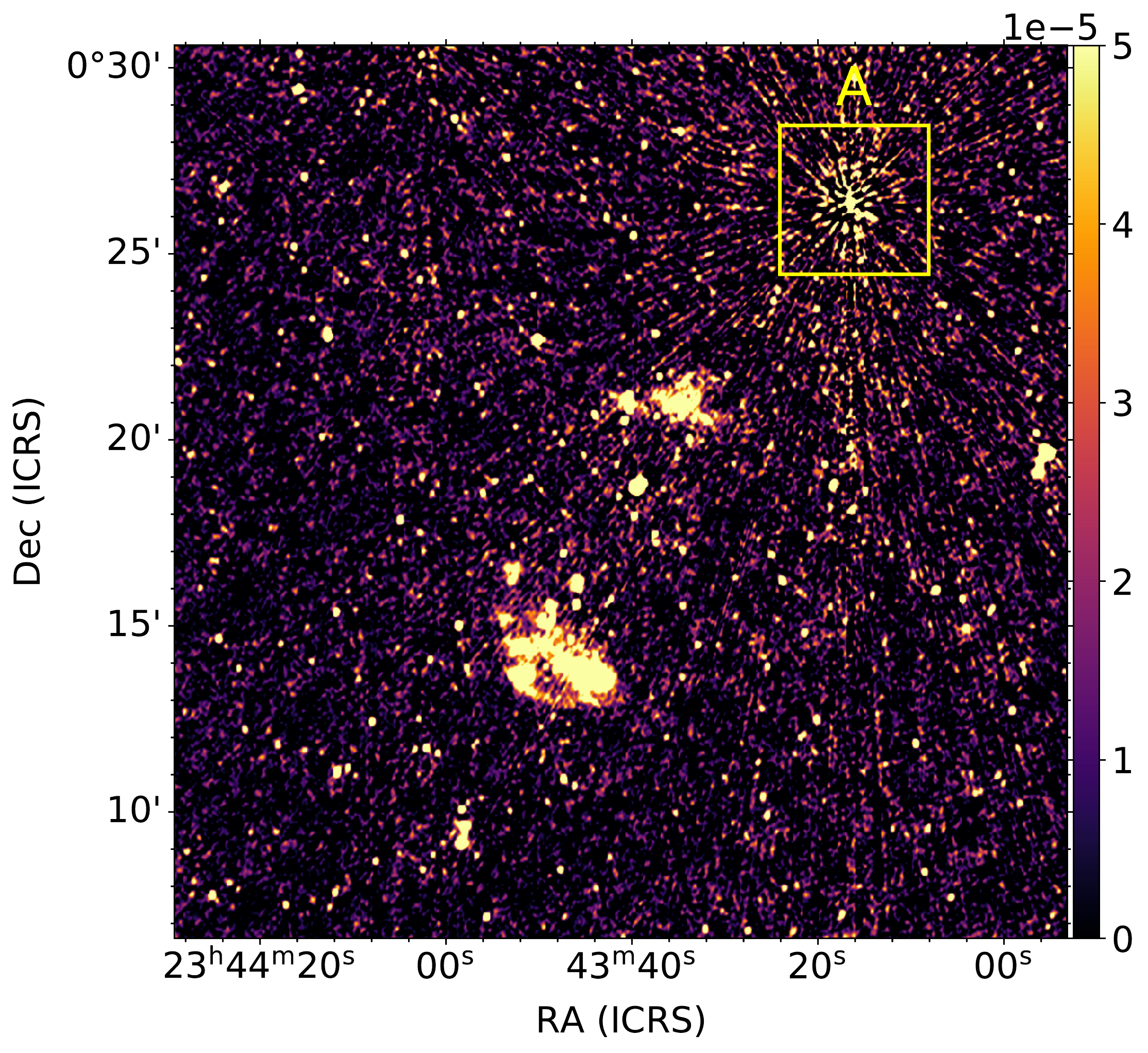}\\
\includegraphics[scale=0.33]{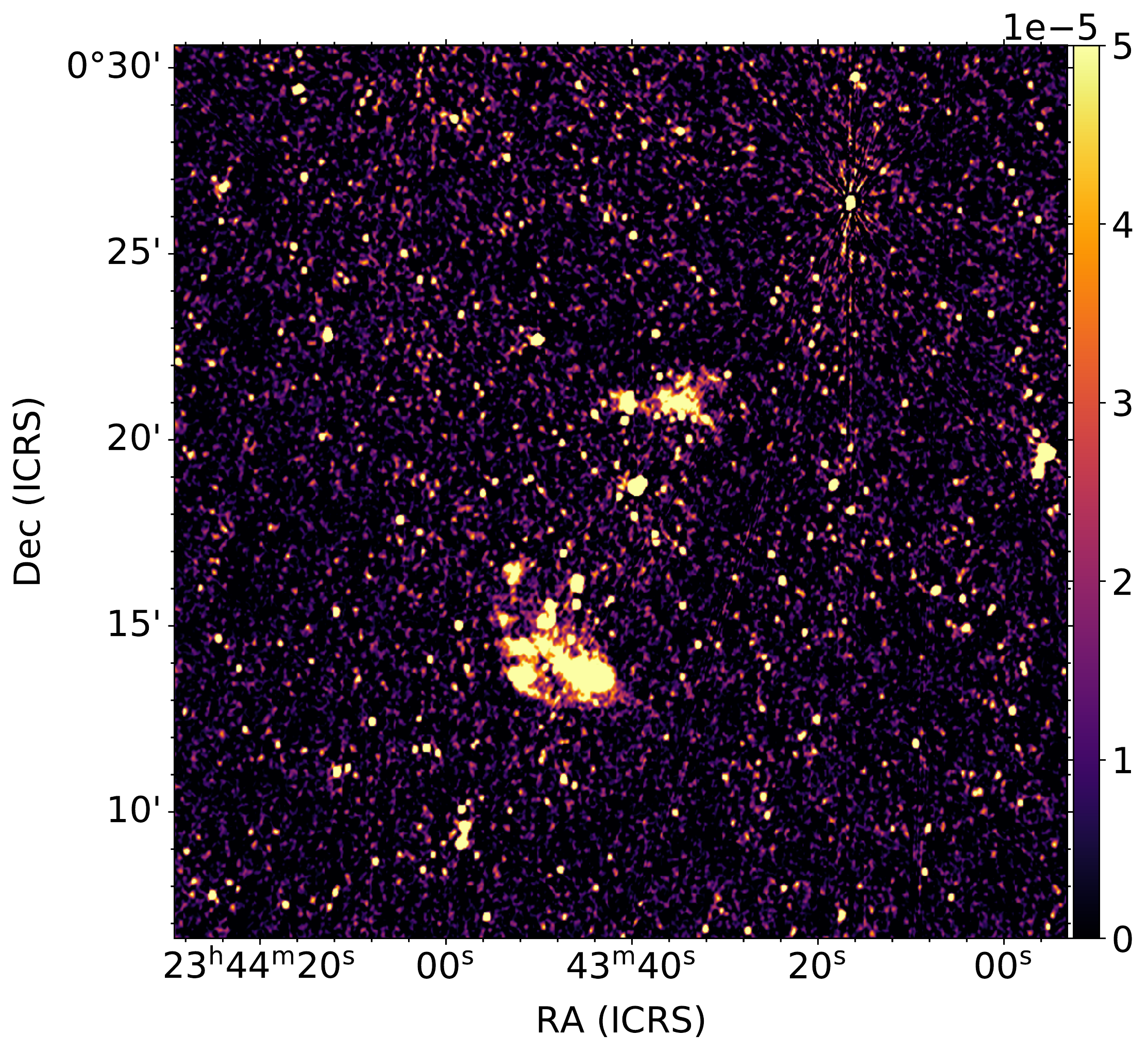}\\
\caption{\textit{Cont}.}
\end{figure}

\begin{figure}[H]\ContinuedFloat
\includegraphics[scale=0.35]{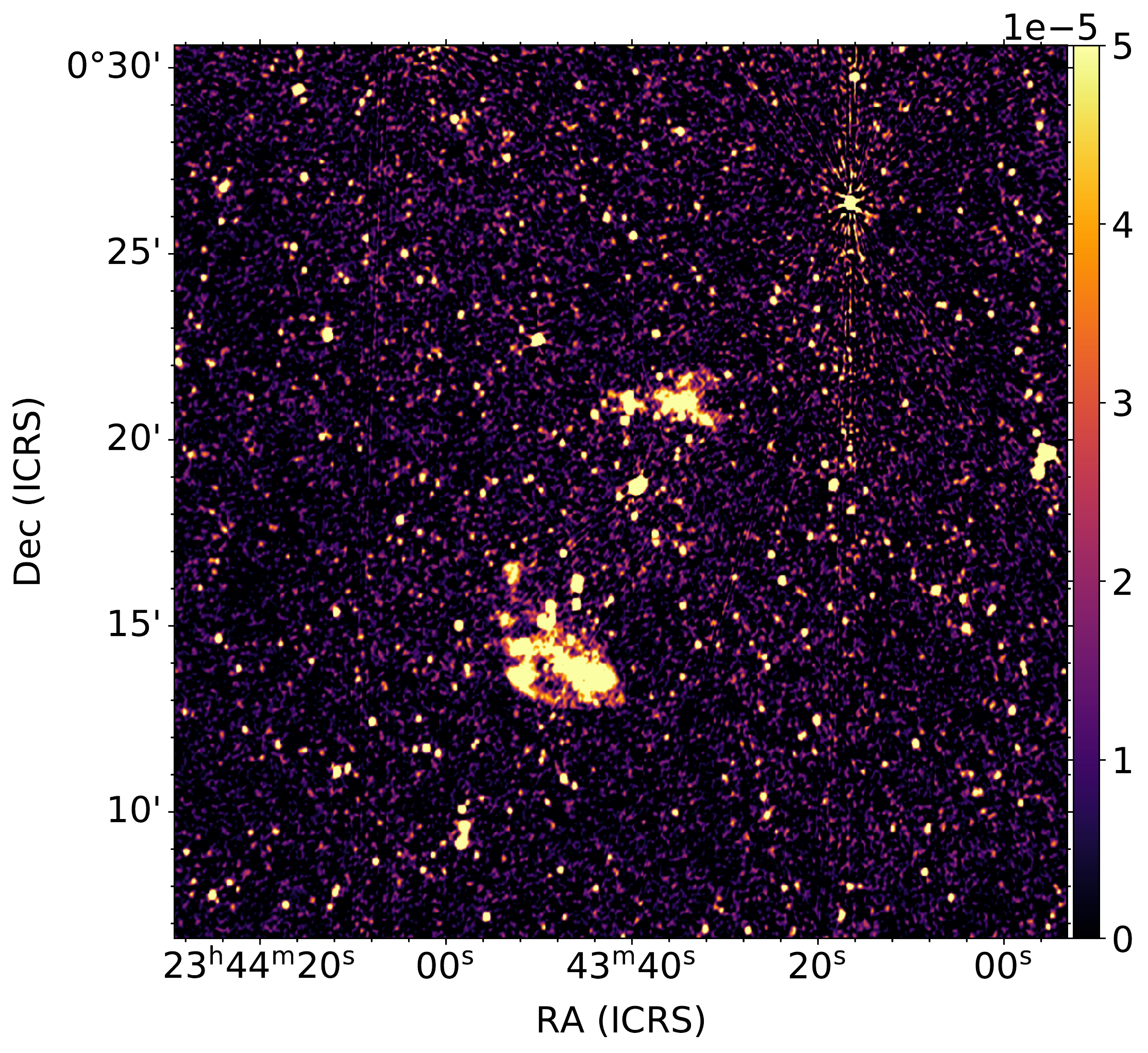}
\caption{{ZwCl 2341.1+0000} central zoomed region where double radio relics are visible. The~order of the image is the same as Figure~\ref{DDE_corr_1}.}
\label{DDE_corr_2}
\end{figure}

\par In Figure~\ref{cutouts_img}, we have shown $8'$ 
 cut-outs of five strong sources (A to E) as mentioned above. From~the top to the bottom row, the~left column is DI images of five sources; the middle column is DDE images generated with the CubiCal and the~right column is DDE images generated with the killMS. We marked every strong source with a green box. We calculated the residual statistics (before and after applying DDEs) in the $8'$ region of every marked source. The~statistical results are listed in Table~\ref{DDE_stats_results}.

\end{paracol}
\nointerlineskip
 \begin{figure}[H]
    \widefigure
     \begin{subfigure}[h]{0.3\textwidth}
         \includegraphics[width=1\textwidth]{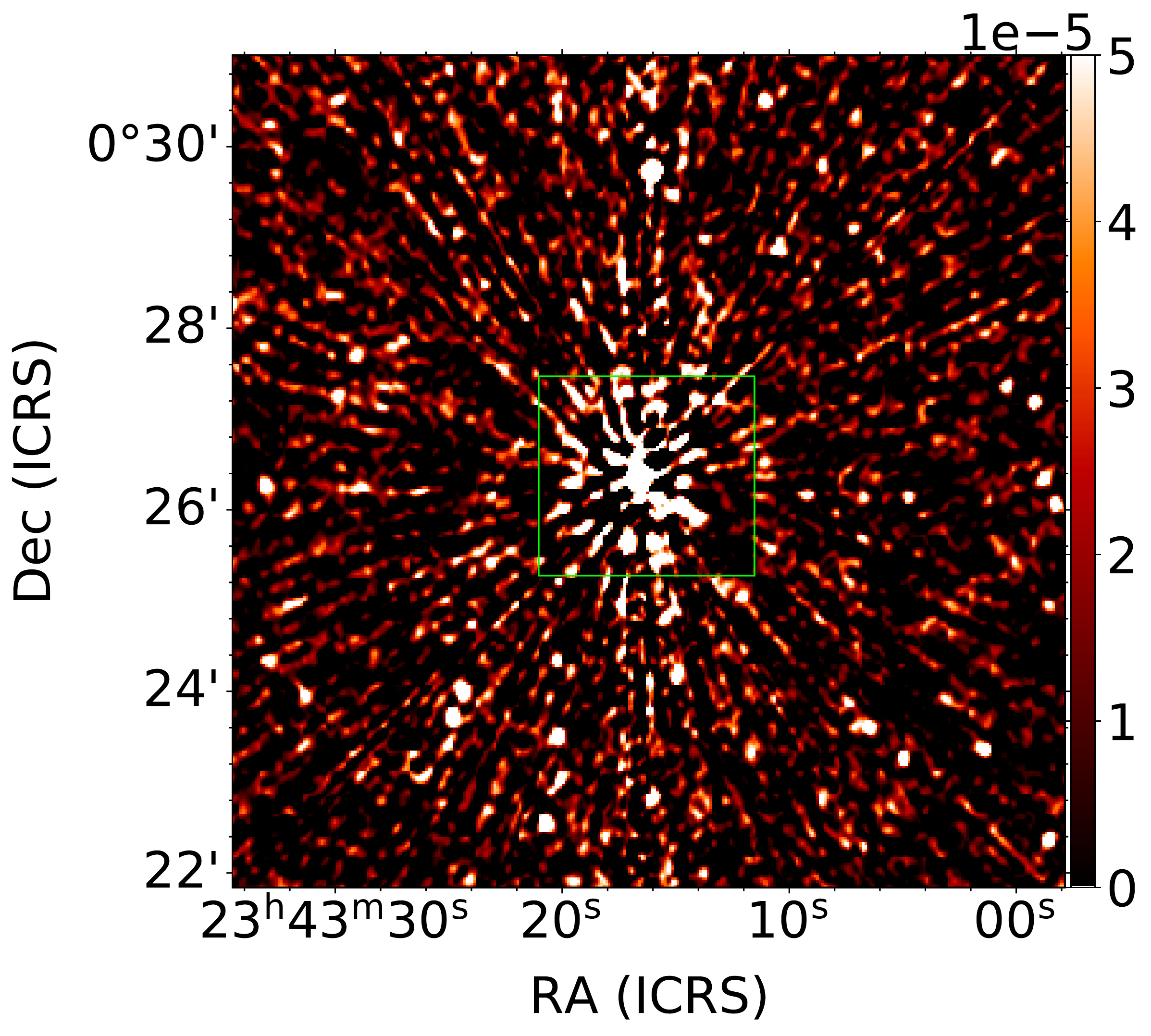}
         
     \end{subfigure}
     \begin{subfigure}[h]{0.3\textwidth}
         \includegraphics[width=1\textwidth]{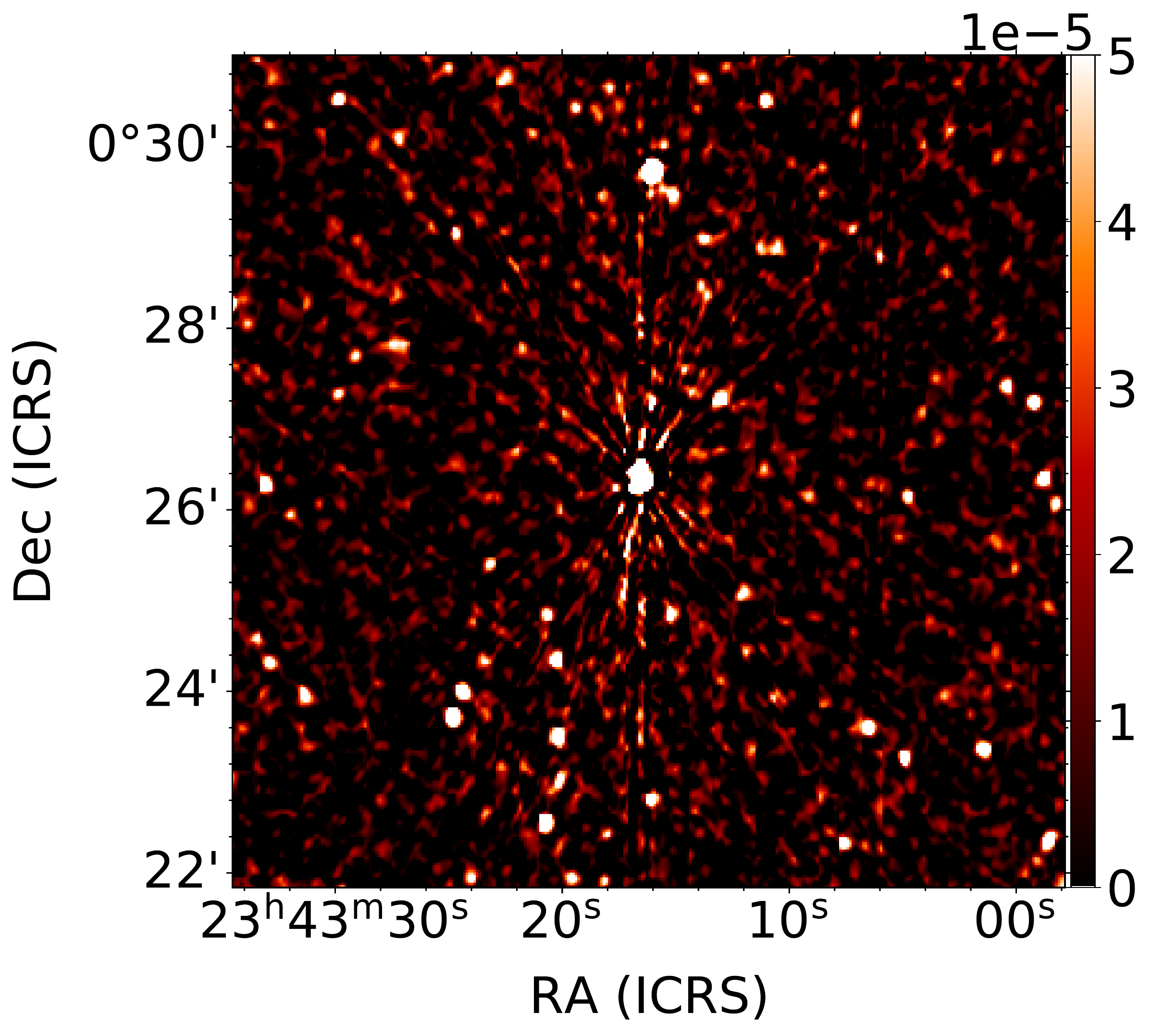}
         
         \end{subfigure}
     \begin{subfigure}[h]{0.3\textwidth}
         \includegraphics[width=1\textwidth]{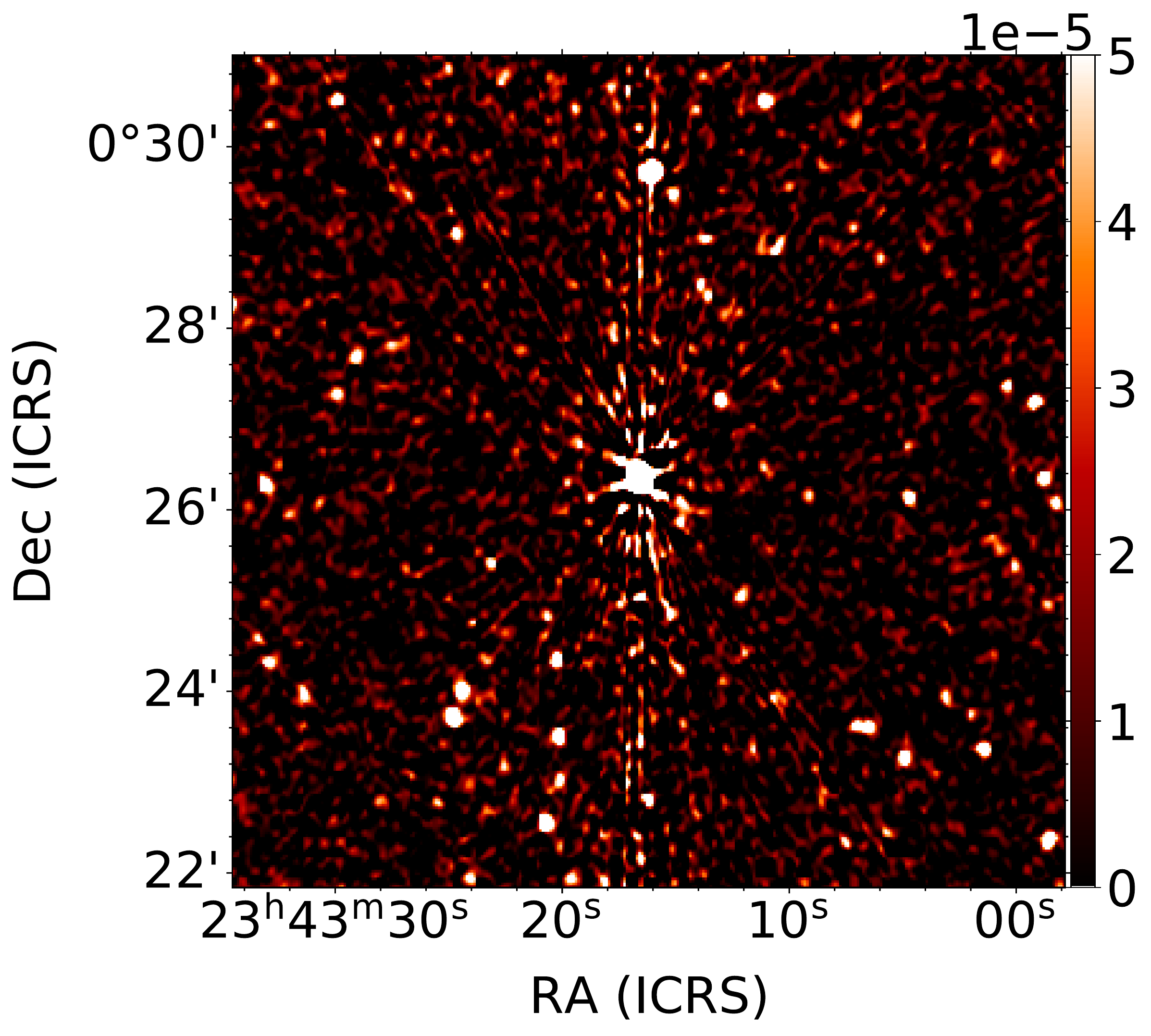}
         
    \end{subfigure}
     \\
    \begin{subfigure}[h]{0.3\textwidth}
         \includegraphics[width=1\textwidth]{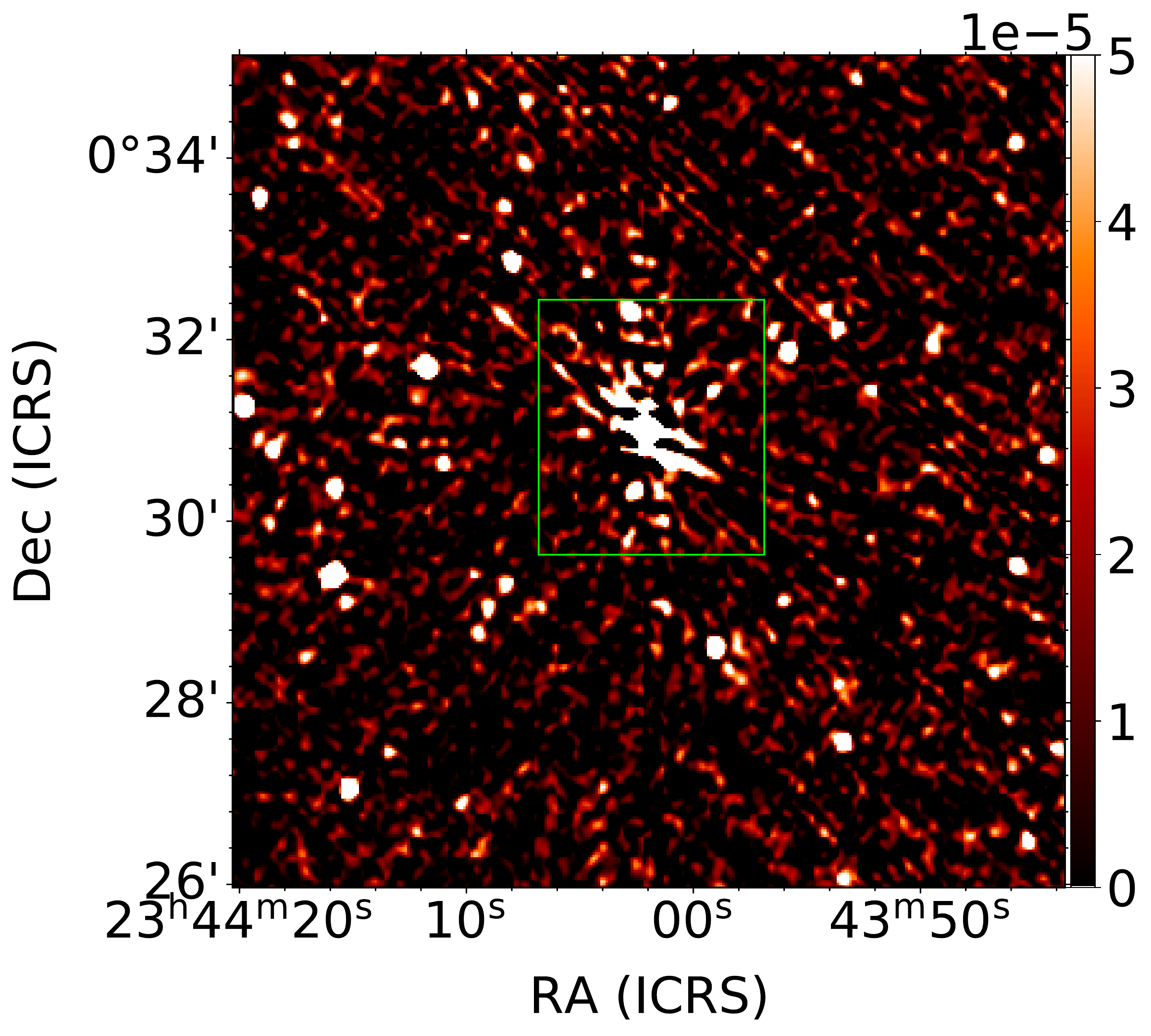}
         
     \end{subfigure}  
      \begin{subfigure}[h]{0.3\textwidth}
         \includegraphics[width=1\textwidth]{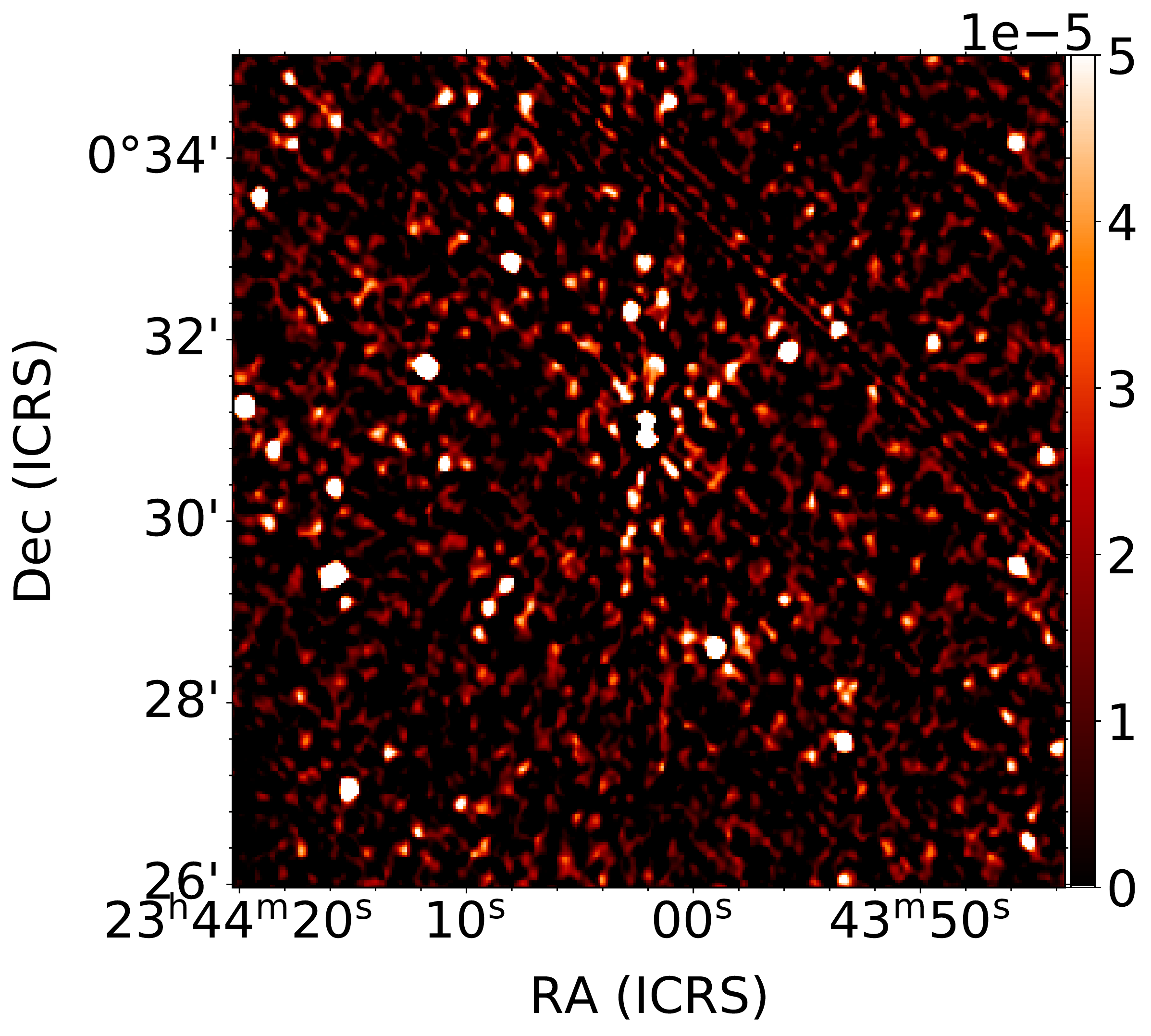}
         
     \end{subfigure}
       \begin{subfigure}[h]{0.3\textwidth}
         \includegraphics[width=1\textwidth]{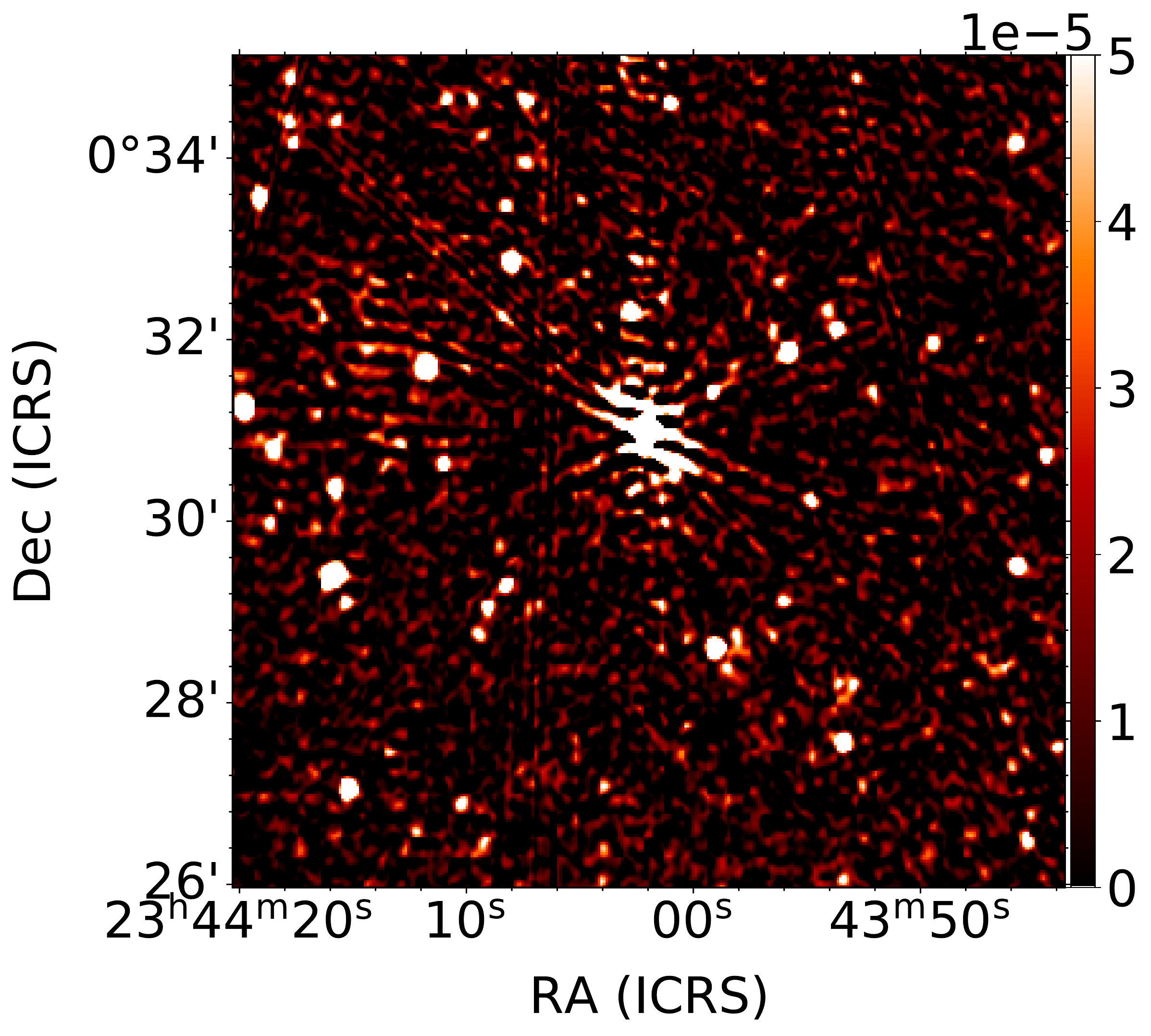}
         
     \end{subfigure}
 
\caption{\textit{Cont}.}
\end{figure}

 \begin{figure}[H]\ContinuedFloat
 \widefigure         
      \begin{subfigure}[h]{0.3\textwidth}
         \includegraphics[width=1\textwidth]{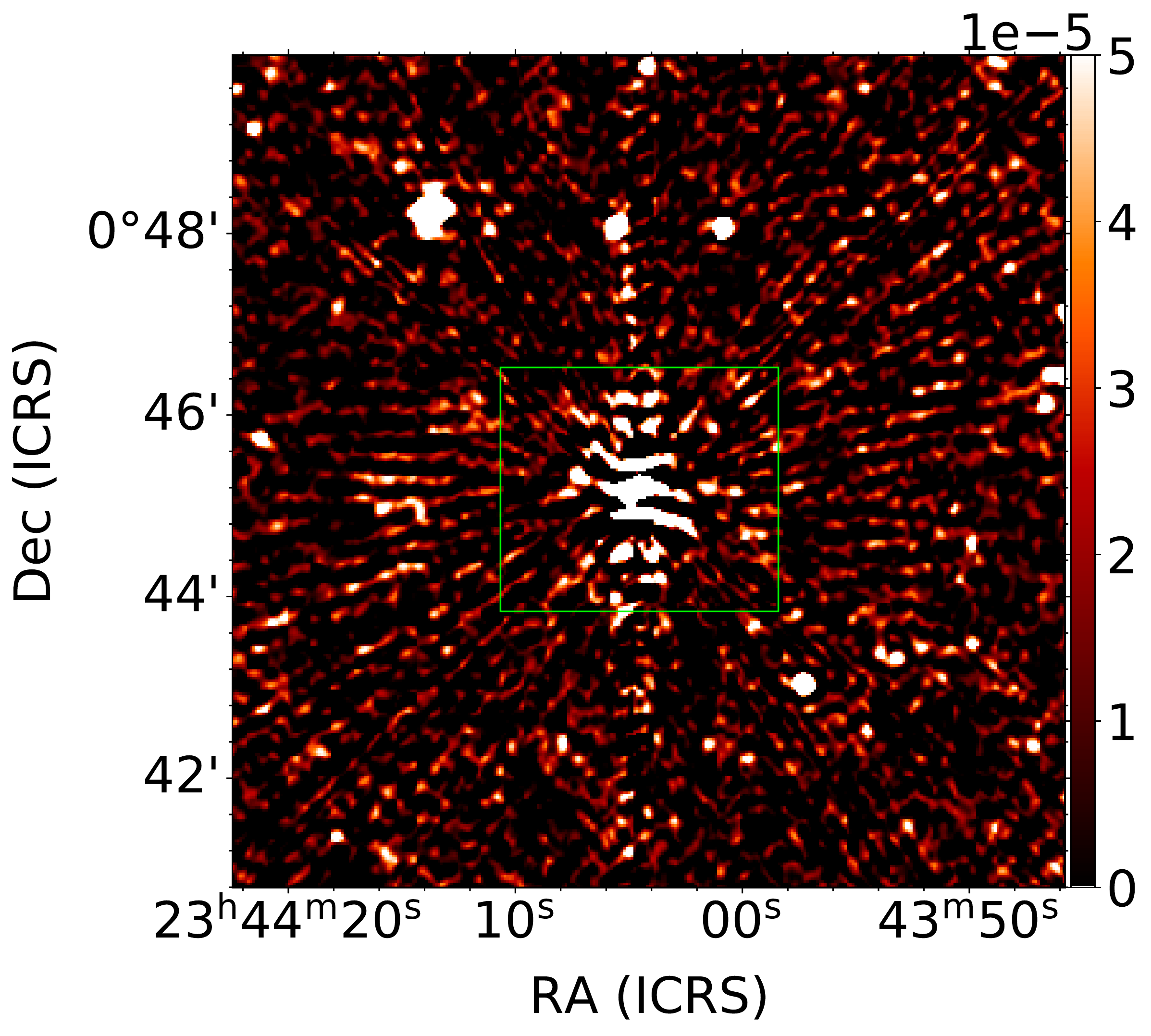}
         
     \end{subfigure}
       \begin{subfigure}[h]{0.3\textwidth}
         \includegraphics[width=1\textwidth]{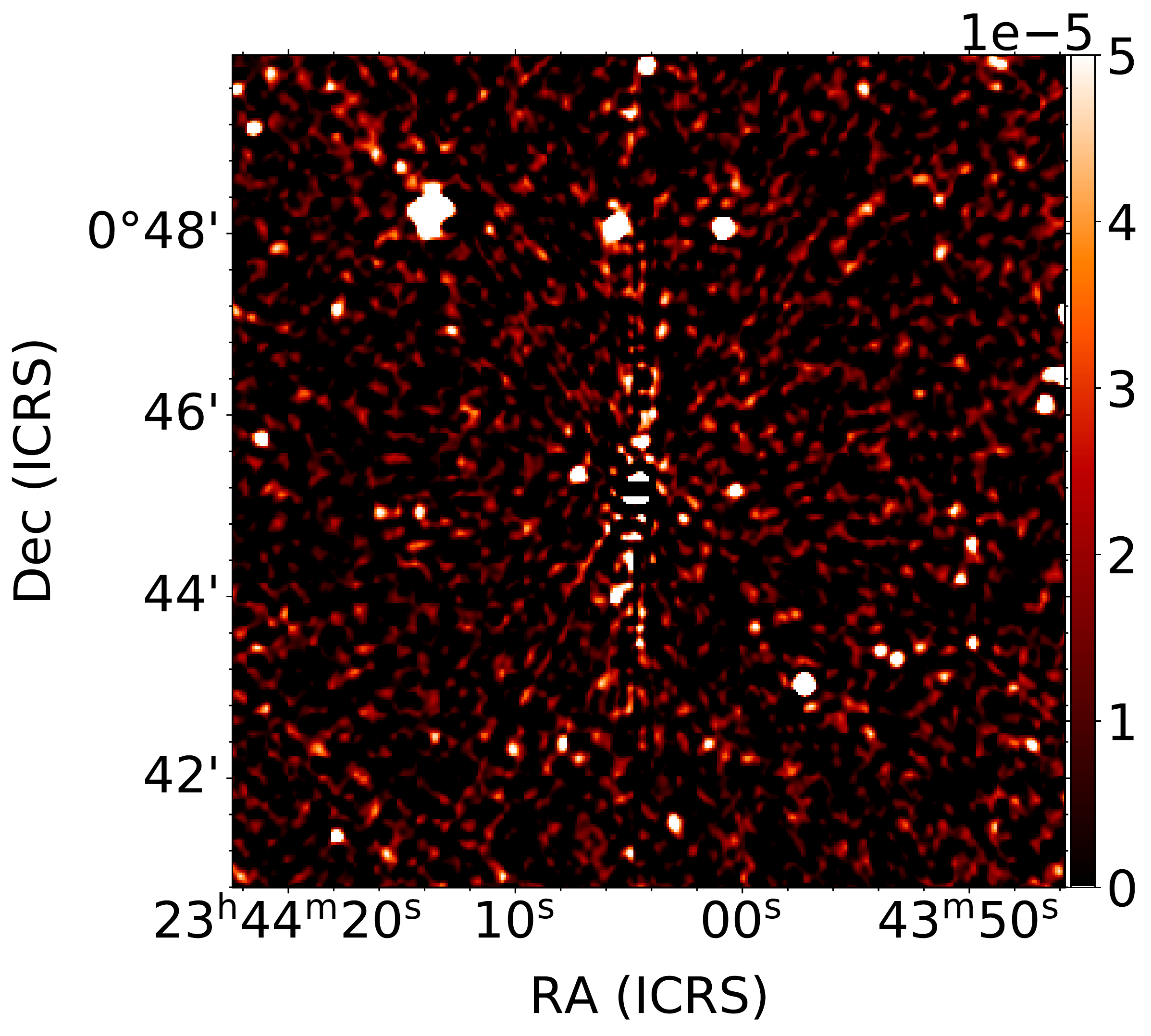}
         
     \end{subfigure}
      \begin{subfigure}[h]{0.3\textwidth}
         \includegraphics[width=1\textwidth]{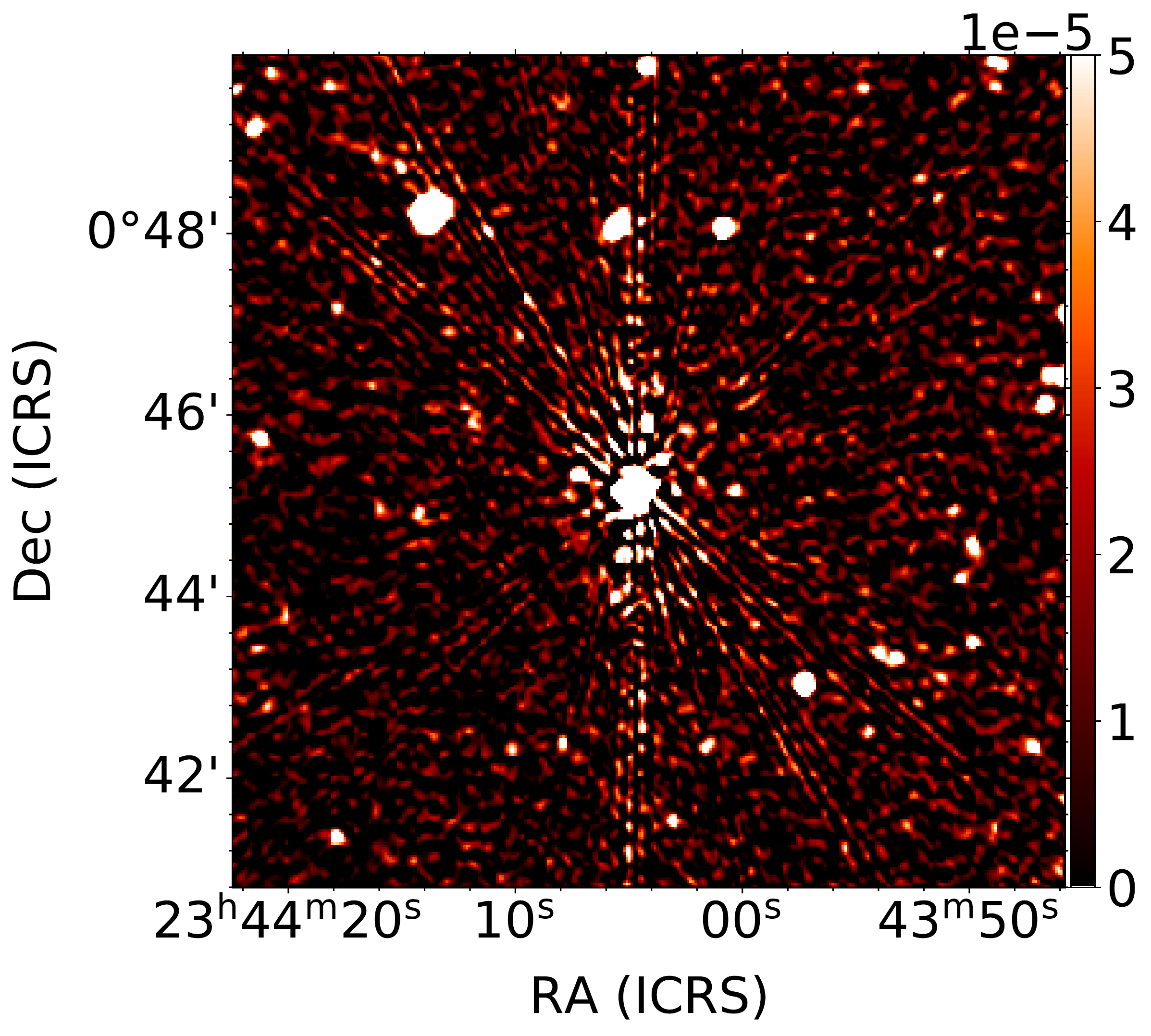}
       
     \end{subfigure}
     \\
   \begin{subfigure}[h]{0.3\textwidth}
         \includegraphics[width=1\textwidth]{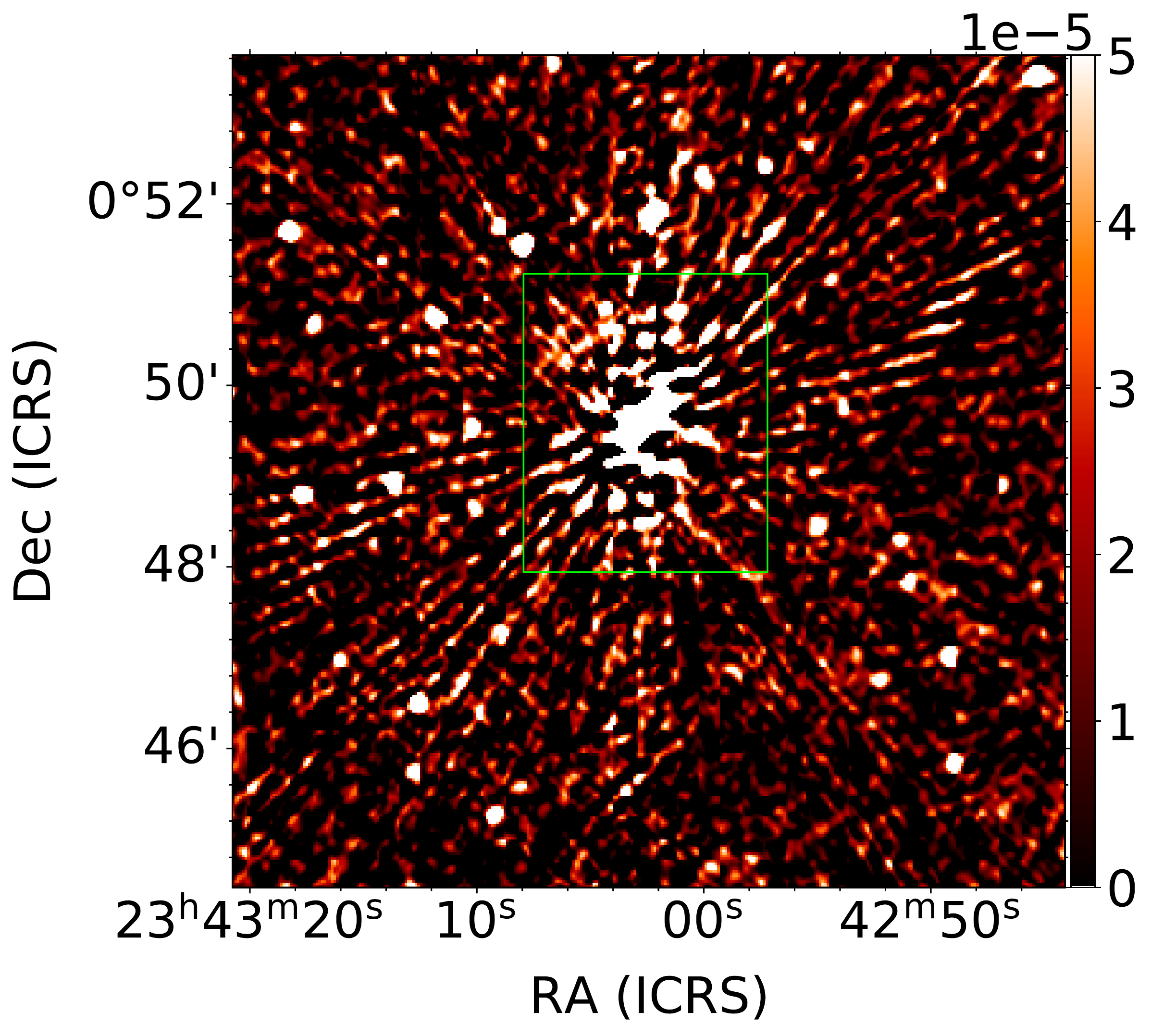}
         
     \end{subfigure}
       \begin{subfigure}[h]{0.3\textwidth}
         \includegraphics[width=1\textwidth]{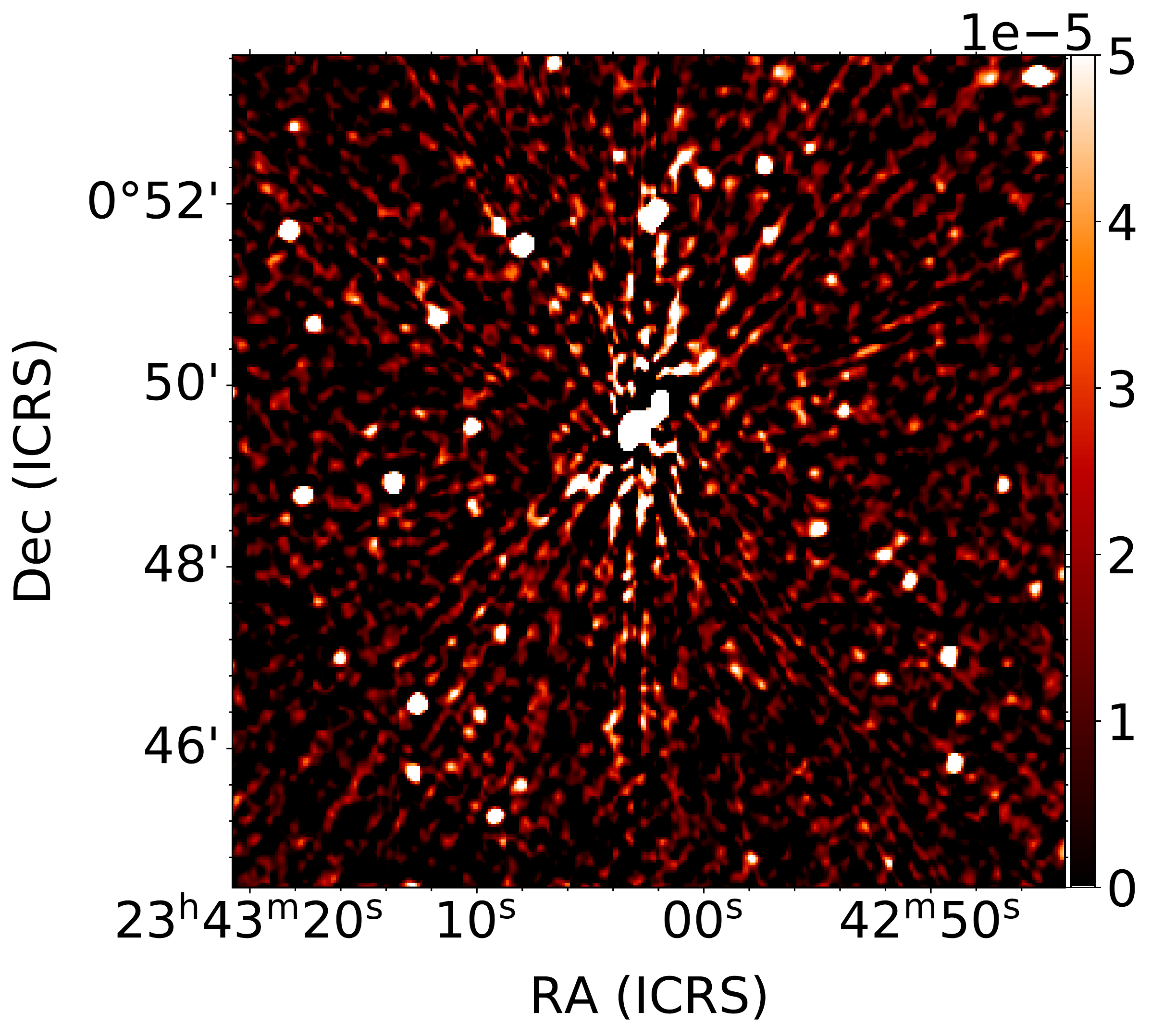}
       
     \end{subfigure}
      \begin{subfigure}[h]{0.3\textwidth}
         \includegraphics[width=1\textwidth]{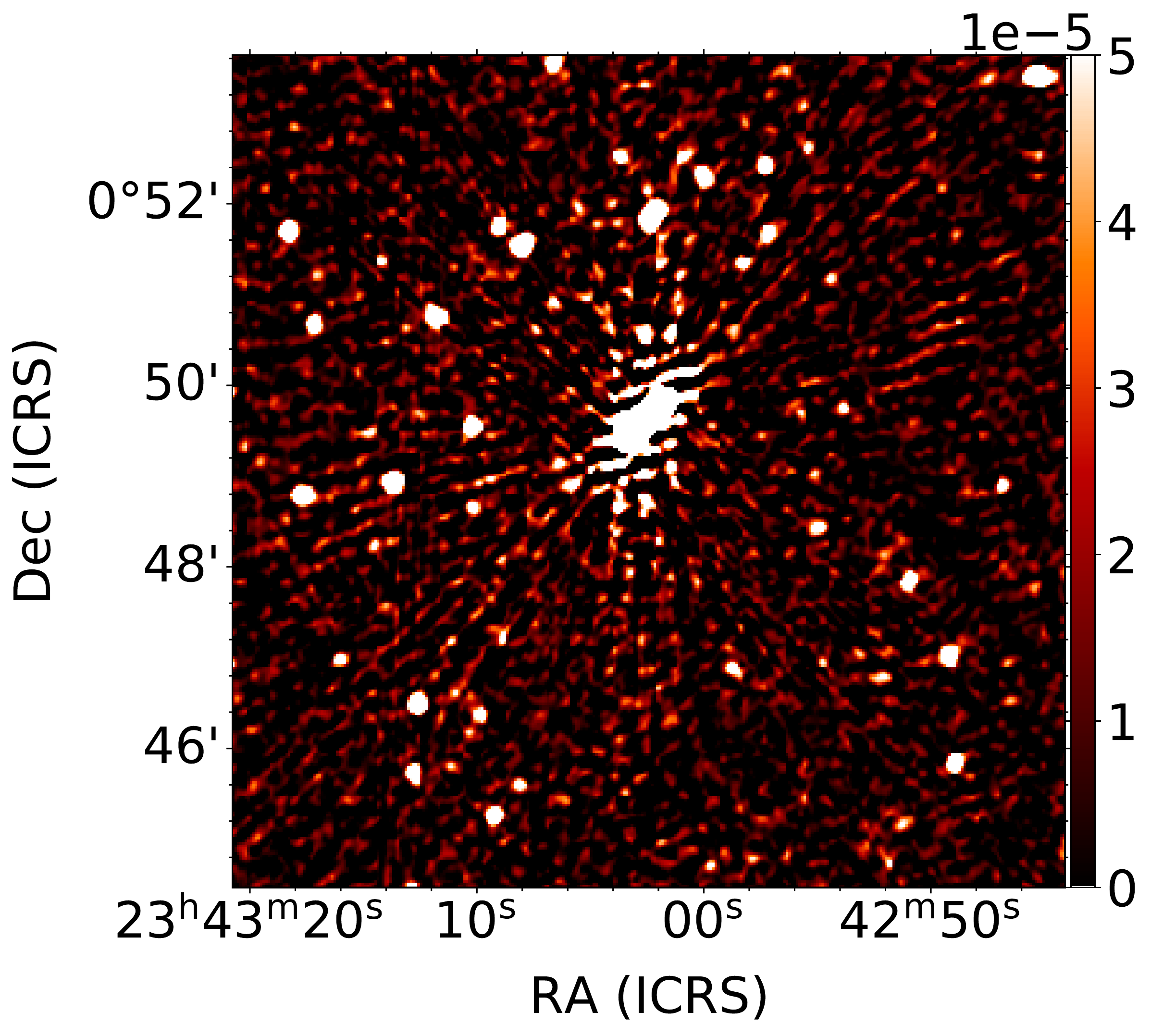}
         
     \end{subfigure}
     \\
           \begin{subfigure}[h]{0.3\textwidth}
         \includegraphics[width=1\textwidth]{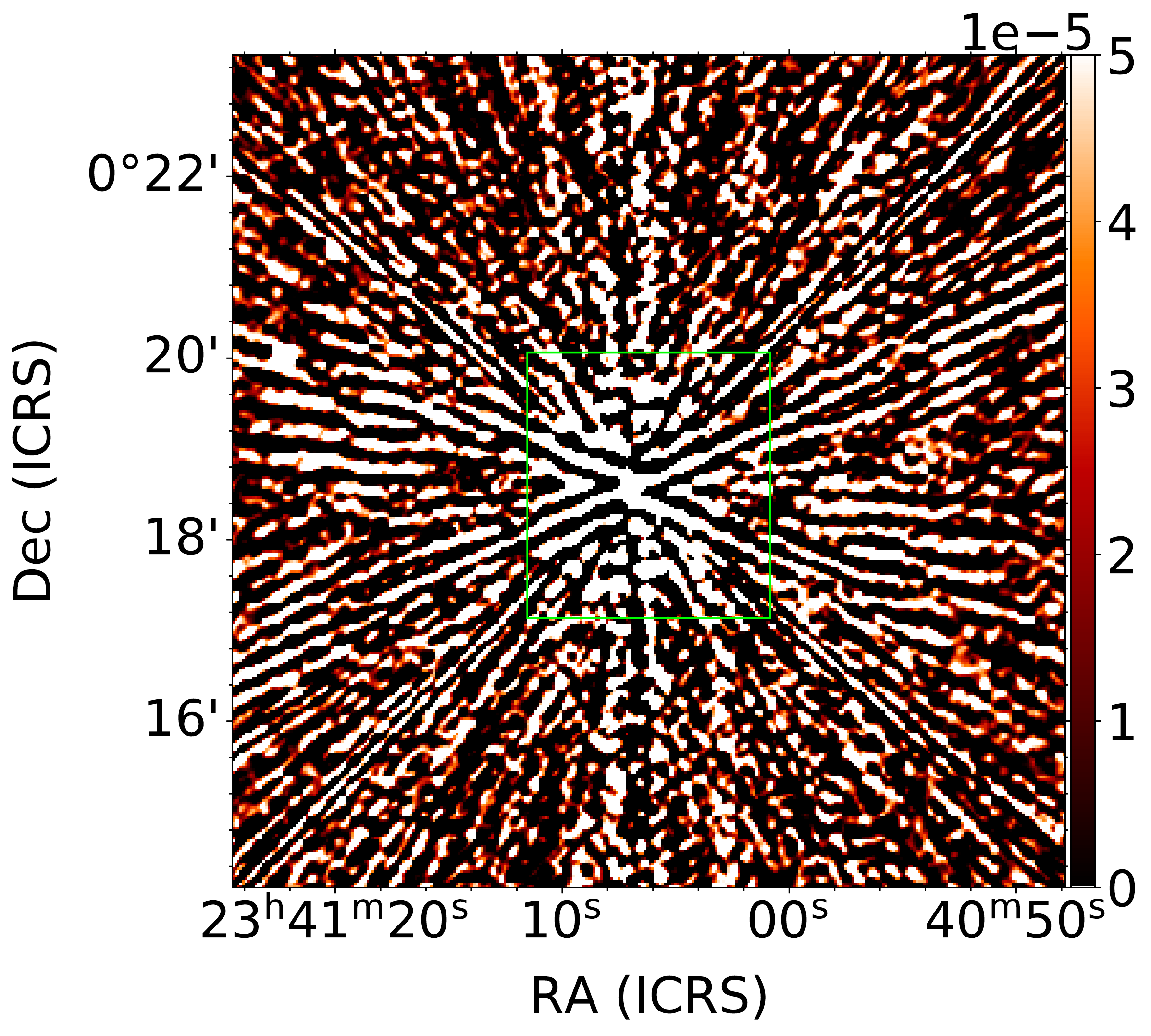}
        
     \end{subfigure}
       \begin{subfigure}[h]{0.3\textwidth}
         \includegraphics[width=1\textwidth]{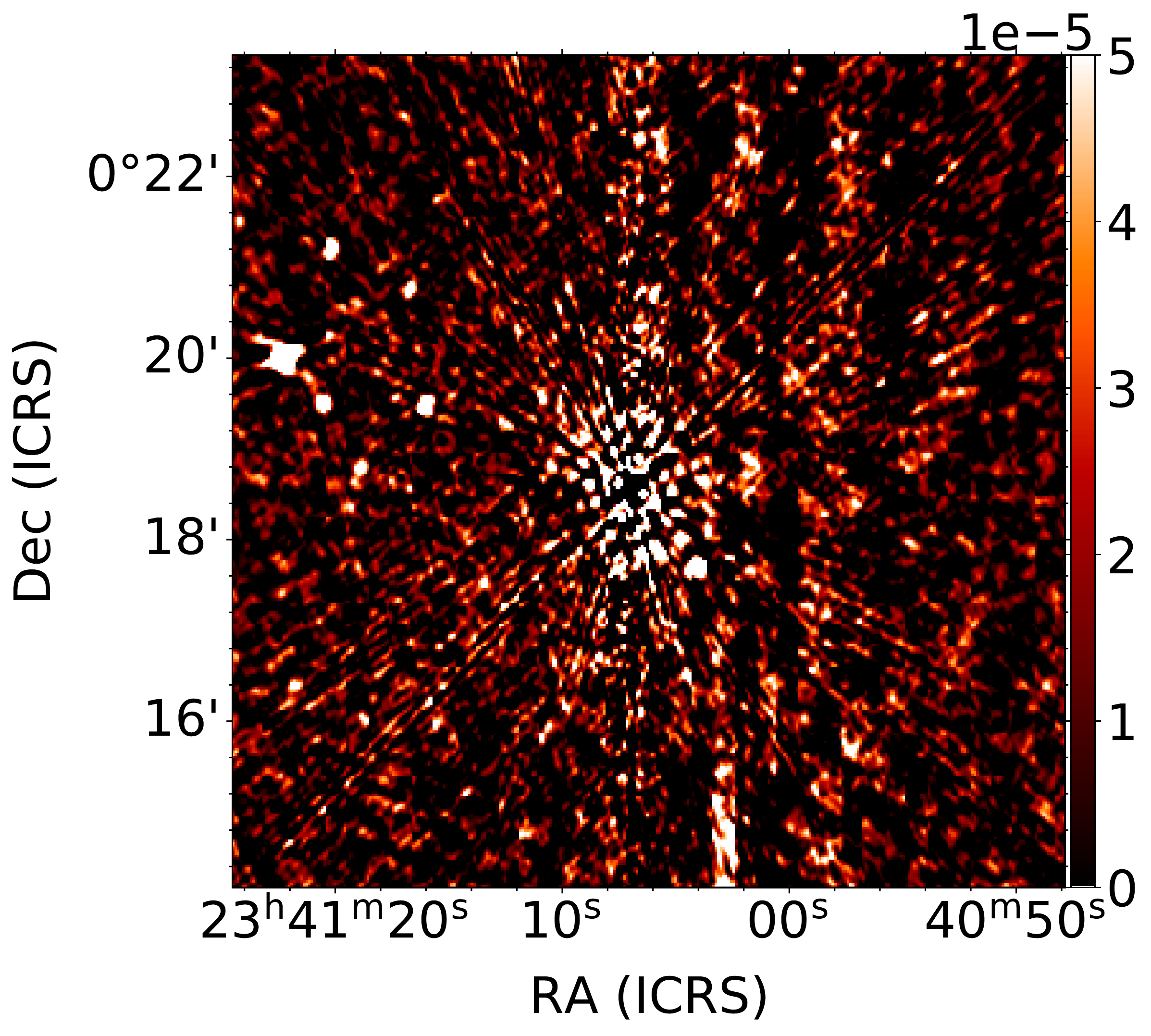}
         
     \end{subfigure}
      \begin{subfigure}[h]{0.3\textwidth}
         \includegraphics[width=1\textwidth]{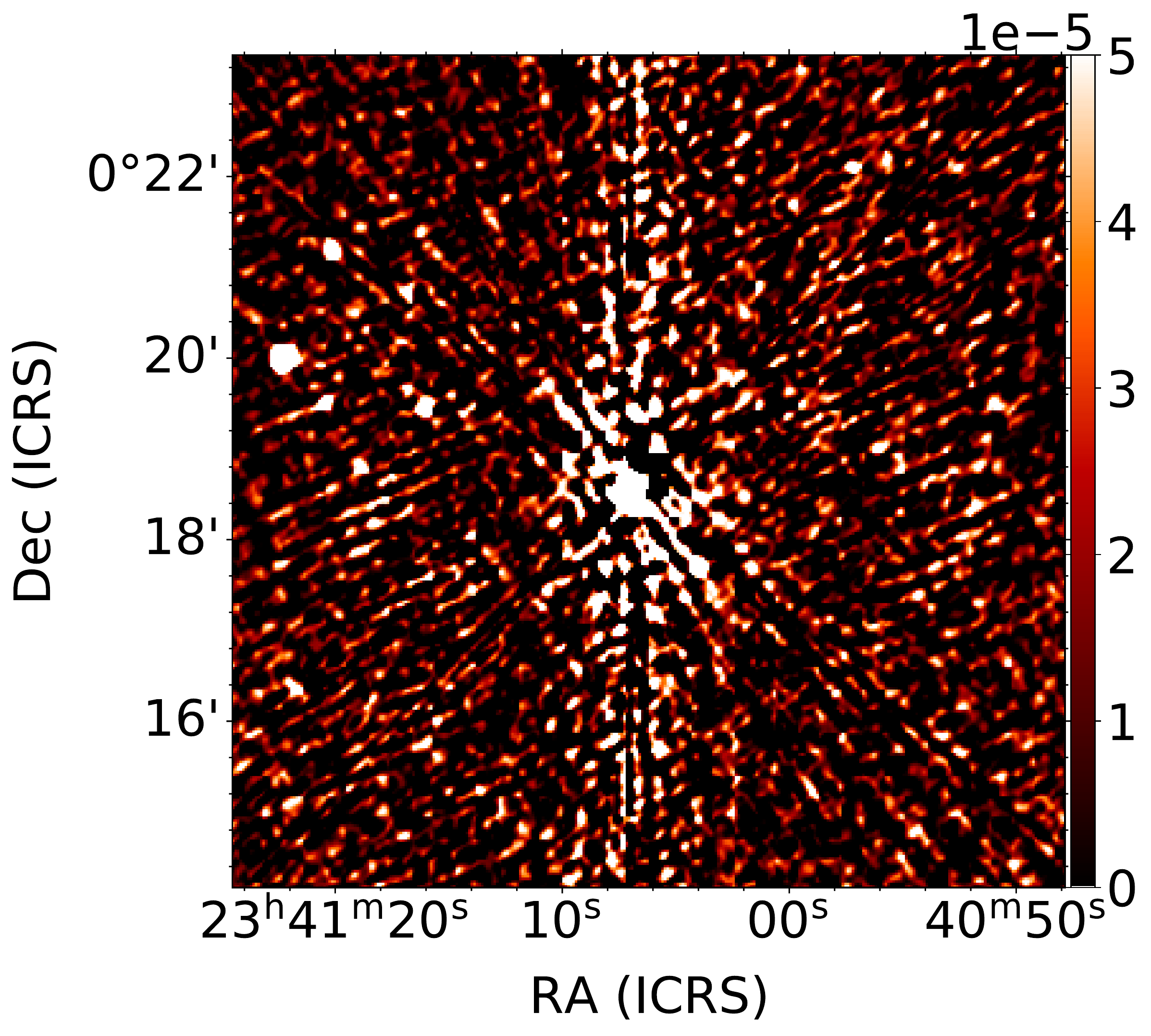}
         
     \end{subfigure}
  \caption{{Cut-outs of sources} A, B, C, D, E from top to bottom: ({\textbf{left}}) column shows DI images, ({\textbf{middle}}) column shows DD images generated with CubiCal and ({\textbf{right}}) column shows DD images generated with~killMS. }
 \label{cutouts_img}
 \end{figure}
\begin{paracol}{2}
\switchcolumn
\vspace{-8pt}
\startlandscape
\begin{specialtable}[H]

\widetable

\caption{Statistics calculated in the residual image, before~and after CubiCal and killMS calibrations, for~five strong sources in the ZwCl 2341.1+0000~field.}\label{DDE_stats_results}

\setlength{\cellWidtha}{\columnwidth/12-2\tabcolsep+0.2in}
\setlength{\cellWidthb}{\columnwidth/12-2\tabcolsep+0in}
\setlength{\cellWidthc}{\columnwidth/12-2\tabcolsep-0.0in}
\setlength{\cellWidthd}{\columnwidth/12-2\tabcolsep-0.0in}
\setlength{\cellWidthe}{\columnwidth/12-2\tabcolsep-0.0in}
\setlength{\cellWidthf}{\columnwidth/12-2\tabcolsep-0.0in}
\setlength{\cellWidthg}{\columnwidth/12-2\tabcolsep-0in}
\setlength{\cellWidthh}{\columnwidth/12-2\tabcolsep-0in}%
\setlength{\cellWidthi}{\columnwidth/12-2\tabcolsep-0.0in}
\setlength{\cellWidthj}{\columnwidth/12-2\tabcolsep-0.1in}
\setlength{\cellWidthk}{\columnwidth/12-2\tabcolsep-0.1in}
\setlength{\cellWidthl}{\columnwidth/12-2\tabcolsep-0in}%
\scalebox{1}[1]{\begin{tabularx}{\columnwidth}{>{\PreserveBackslash\centering}m{\cellWidtha}>{\PreserveBackslash\centering}m{\cellWidthb}>{\PreserveBackslash\centering}m{\cellWidthc}>{\PreserveBackslash\centering}m{\cellWidthd}>{\PreserveBackslash\centering}m{\cellWidthe}>{\PreserveBackslash\centering}m{\cellWidthf}>{\PreserveBackslash\centering}m{\cellWidthg}>{\PreserveBackslash\centering}m{\cellWidthh}>{\PreserveBackslash\centering}m{\cellWidthi}>{\PreserveBackslash\centering}m{\cellWidthj}>{\PreserveBackslash\centering}m{\cellWidthk}>{\PreserveBackslash\centering}m{\cellWidthl}}
\toprule

\multirow{2}{*}{\textbf{Source} } &    \textbf{RMS}   &  \multirow{2}{*}{\textbf{ DR1}} &\multirow{2}{*}{\textbf{DR2 }}&\multirow{2}{*}{\textbf{DR3 }} &   \textbf{MAD}  &   \textbf{MIN}   &  \textbf{MAX}  &  \textbf{SUM NEG}  &  \multirow{2}{*}{\textbf{SKEW}}  &  \multirow{2}{*}{\textbf{KURT}}  &  \multirow{2}{*}{\textbf{NORM} } \\
 &    \boldmath\textbf{$\upmu$Jy beam$^{-1}$}  &  & &  & \boldmath \boldmath\textbf{$\upmu$Jy beam$^{-1}$}  &  \boldmath\textbf{mJy beam$^{-1}$}  & \boldmath \textbf{mJy beam$^{-1}$}  &  \boldmath\textbf{mJy beam$^{-1}$ } &    &   & \\
\midrule
A  &      &    &     &     &    &    &    &    &   & & \\
\midrule
DI  &  37.0 & 2288.95 & 66.77  & 4282.69 & 15.0 & $-$1.27 & 84.61  & $-$1170.39 & $-$6.59 & 175.62 & 132287.84 \\
DD (CubiCal)  &  16.0 & 5384.67 & 521.17 & 6011.81 & 9.0  & $-$0.16 & 1.93   & $-$642.60  &  1.2  & 12.53  & 32982.4 \\
DD (killMS) &  18.0 & 4819.0 & 207.25 & 6728.32 & 8.0  & $-$0.41 & 84.35  & $-$648.68  & $-$2.19 & 61.92  & 68730.32  
\\
\midrule
B  &      &    &     &     &    &    &    &    &   & & \\
\midrule
DI  &  23.0 & 1452.71 & 89.54  & 1703.88 & 11.0 & $-$0.38 & 33.66 & $-$780.03  & 1.26 & 41.87 & 48327.62 \\
DD  &  17.0 & 1948.45 & 203.95 & 2391.82 & 9.0  & $-$0.17 & 1.11  & $-$640.55  & 1.53 & 13.83 & 40406.3 \\
DD  &  19.0 & 1729.23 & 89.72  & 2616.98 & 8.0  & $-$0.37 & 32.80 & $-$525.23  & 0.78 & 63.84 & 43861.46 
\\
\midrule
C  &      &    &     &     &    &    &    &    &  & &  \\
\midrule
DI  &  37.0 & 859.03 & 35.10  & 1587.62 & 12.0 & $-$0.89 & 31.37 & $-$974.88  & 2.47  & 161.56 & 81404.9  \\
DD  &  18.0 & 1786.54 & 123.52 & 2228.62 & 9.0  & $-$0.25 & 3.19  & $-$683.07  & 0.56  & 23.3   & 29623.6  \\
DD  &  23.0 & 1329.44 & 39.44  & 2416.91 & 9.0  & $-$0.77 & 30.30 & $-$647.93  & $-$6.74 & 207.55 & 134959.27 
\\
\midrule
D  &      &    &     &     &    &    &    &    &  & &  \\
\midrule
DI  &  38.0 & 823.34 & 39.83 & 1596.63 & 15.0 & $-$0.79 & 31.54 & $-$1009.09 & $-$1.06 & 91.05  & 52000.85  \\
DD  &  24.0 & 1335.30 & 57.60 & 2241.26 & 10.0 & $-$0.55 & 2.09  & $-$668.20  & $-$3.96 & 86.19  & 97598.32  \\
DD  &  23.0 & 1293.42 & 52.52 & 2414.77 & 9.0  & $-$0.58 & 30.27 & $-$672.90  & $-$4.67 & 102.88 & 107886.23 
\\
\midrule
E  &      &    &     &     &    &    &    &    &   & & \\
\midrule
DI  &  201.0 & 903.81  & 31.46  & 9191.59 & 47.0 & $-$5.77 & 181.59 & $-$4077.27 & $-$5.1  & 196.17 & 117774.02 \\
DD  &  30.0  & 6097.83 & 237.53 & 12902.67 & 14.0 & $-$0.76 & 2.60   & $-$895.05  & 0.29  & 76.61  & 38640.5 \\
DD  &  36.0  & 4446.85  & 233.40 & 12927.69 & 17.0 & $-$0.69 & 162.08 & $-$1178.32 & $-$1.13 & 33.72  & 43613.94  
\\
\midrule
Full image  &      &    &     &     &    &    &    &    &   &   &  \\
\midrule
DI  & 20.0 & - & 31.46  & 9191.59  & 9.0 & $-$5.77 & 181.59 & $-$196953.75   & $-$15.45 & 5995.31 & 75889133.95 \\
DD  & 14.0 & - & 115.12 & 12902.67 & 7.0 & $-$1.58 & 42.21 & $-$154531.51   & $-$1.59 & 636.36 & 26984110.13\\
DD  & 12.0 &- & 122.20 & 12927.69 & 7.0 & $-$1.33 & 162.08 & $-$154750.56   & $-$4.86 & 531.9 & 43647231.04\\ \bottomrule

\end{tabularx}}
\end{specialtable}
\finishlandscape

\section{Discussion and~Conclusions} \label{Discussion}
\par In this work, our primary goal was to test both DDE methods (CubiCal and killMS) independently on wide-band MeerKAT data in order to the improve image quality and quantify it statistically. As~a quantitative measure, we calculated the rms, minimum pixel value (Min), maximum pixel value (Max), DR, sum of the negative pixels value (Sum), mean absolute deviation (MAD), skewness (Skew), kurtosis (Kurt) and normality test (Norm) values in an 8$''$  $\times$ 8$''$ box centred on each bright source. In~these calculations, we used both residual and restored maps generated by the DDFacet. We calculated the statistics for the cut-outs of five strong sources as well as full images of DI and DD. In~this work, our aim was not to compare two distinct DDE techniques (CubiCal and killMS), as~both work differently and have different calibration techniques. The~calibration process also depends on their input parameter selections, underlying data qualities (depending on the level of background noise present in the data, both techniques apply calibration differently) and the purpose of the scientific case. 

\par Our approach of statistical and quantitative analysis is useful with regards to studying image fidelity and DR. Image fidelity is an important indicator of how well the sky or source brightness distribution is accurately regenerated in an image, while DR is a measure of the degree to which imaging artefacts or calibration errors around the strong source(s) are suppressed, which imply a higher fidelity of the on-source reconstruction. The~DR of an image is generally measured by the ratio of the maximum intensity to the image rms (local or global). For~five strong sources, within the~cut-out regions, we calculated DR by taking the ratio between (1) Max pixel and the local rms (DR1 = $\frac{\mathrm flux_{\mathrm peak}}{\mathrm rms_{\mathrm local}}$), (2) Max pixel and the absolute Min (DR2 = $\frac{\mathrm flux_{\mathrm peak}}{|{\mathrm min_{\mathrm local}|}}$), and~(3) Max pixel and the global rms (DR3 = $\frac{\mathrm flux_{\mathrm peak}}{\mathrm rms_{\mathrm global}}$). For~calculating the Max pixel values, we used restored images. CubiCal peels sources, hence reducing their peak flux densities in restored images. Therefore, to~calculate the DR for a CubiCal DD image, we used the Max pixel value of corresponding DI image. The~other statistical parameters (rms, Min, Sum, MAD, Skew, Kurt and Norm) are calculated from the residual map. In~this work, the~motive for including negative pixel statistics is that imaging artefacts generated by calibration errors manifest themselves in a positive-negative pattern (since psf sidelobes can be positive/negative), while the true astrophysical source emission is all positive. The~distribution of the negative pixels is therefore suggestive of the number of artefacts. 
\par As we can see in Table~\ref{DDE_stats_results}, there are clear decrements in the (absolute of) sum of negative and Min values after DDE calibration. The local rms and MAD values are also improved after the DDE in both techniques. As~both of these parameters (rms and MAD) measure the deviation from the mean, they calculate the same quantities. Due to rms decreasing after DDE, the DR values are higher for DD images, which implies clear improvements around the strong sources those are causing artefacts. We note that for five cut-outs, the~DR1 and DR2 are better for CubiCal-generated DD images as compared to the killMS, while DR3 is better for killMS. Furthermore, for~full images, the~DR2 and DR3 are better for killMS. This is because CubiCal performs peeling for strong individual sources while killMS corrects the full image. This suggests CubiCal gives better statistics for the local region while killMS gives better statistics for the global region. {{For our future work; to study radio galaxy} (compact sources) properties and their optical counterparts, we will use the killMS-generated DD image. The~response of DD calibration on faint and extended radio emission is a subject of further research. It was noticed that the unmodelled radio flux of extended source is absorbed by the calibration process \citep{2021A&A...648A...1T}.} 

\par For the other three parameters, Skew, Kurt and Norm, we could not find a clear separation between DI and DD (for both CubiCal and killMS) and they have limited use in our work. These three parameters basically trace the symmetries in the noise or how much deviation exists from the normal distribution, and~outliers in the given data. The~killMS DD images have more negative Skew values than CubiCal. For a~full image, the~Skew values are better for DD than DI. The~z-score values are too large for Kurt and Norm tests, indicating the absence of Gaussian distributions in the background noise. For~the Norm test, the~probabilities (we have not shown $p$-values in Table~\ref{DDE_stats_results}) are less than the alpha value (0.05), so the null hypothesis (samples come from a normal distribution) can be rejected. In the radio image, residual noise is correlated with frequency-dependent psf and there is little or no effects of DDE on the background noise distribution. We have also compared these trends in restored images, and~the results are the same.
\par In this paper, we have shown the importance of DDE for new-generation radio telescopes. We have applied two different DDE methods, CubiCal and killMS, to~the MeerKAT data of the {Saraswati} supercluster and shown the improvements in image quality to better study diffuse radio sources. We also calculated several statistical parameters to quantify the improvements. In~this analysis, we found the sum of negative, Min, rms and MAD values are clearly showing the improvements around strong sources. There is a clear increment in the DR around these strong sources.

\vspace{6pt}

\authorcontributions{Formal analysis, investigation and~writing---original draft preparation---V.P. and R.K.; Writing---review and editing---B.H., A.R. and N.O. All authors have read and agreed to the published version of the~manuscript.}

\funding{The financial assistance of the South African Radio Astronomy Observatory (SARAO) towards this research is hereby acknowledged.  This work is based upon research supported by the South African Research Chairs Initiative of the Department of Science and Technology and National Research Foundation. }
\dataavailability{The data presented in this study are openly available in SARAO archival system [\url{https://archive.sarao.ac.za/}].}

\acknowledgments{We thank Oleg Smirnov and Cyril Tasse for their fruitful discussion. The~MeerKAT telescope is operated by the South African Radio Astronomy Observatory, which is a facility of the National Research Foundation, an~agency of the Department of Science and~Innovation.}

\begin{adjustwidth}{-4.6cm}{0cm}
\printendnotes[custom]
\end{adjustwidth}

\end{paracol}
\reftitle{References}
\externalbibliography{yes}

\end{document}